\documentclass[a4paper,11pt]{article}
\pdfoutput=1 

\usepackage{jheppubv2} 

\usepackage[T1]{fontenc} 
\usepackage{tikz}
\usepackage{tikz-network}
\usepackage{xcolor}
\usepackage{mathrsfs, amssymb, amsbsy}
\usepackage{multirow}

\usetikzlibrary{patterns}
\usetikzlibrary{matrix,positioning,fit}
\usetikzlibrary{calc}
\newdimen\R
\def\Tr{\mathop{\mbox{\normalfont Tr}}\nolimits}
\def\chartheta#1#2{\Theta\!\left[\begin{smallmatrix}#1\\#2\end{smallmatrix}\right]\!}

\usepackage{dcolumn}
\newcolumntype{L}{D{.}{.}{2,8}}

\newcommand{\beq}{\begin{equation}}
\newcommand{\eeq}{\end{equation}}

\title{\boldmath  Crossing-symmetric Twist Field Correlators and Entanglement Negativity in Minimal CFTs}

\author[a, b]{Filiberto Ares,}
\author[c]{Raoul Santachiara,}
\author[d, e]{Jacopo Viti}

\affiliation[a]{International Institute of Physics, UFRN, \\ Campos Universit\'ario, Lagoa Nova 59078-970 Natal, Brazil}
\affiliation[b]{International School for Advanced Studies (SISSA), \\ Via Bonomea 265, 34136 Trieste, Italy}
\affiliation[c]{Universit\'e Paris-Saclay,  CNRS,  LPTMS,  \\ 91405,  Orsay,  France}
\affiliation[d]{International Institute of Physics \& ECT, UFRN, \\ Campos Universit\'ario, Lagoa Nova 59078-970 Natal, Brazil}
\affiliation[e]{INFN, Sezione di Firenze, \\ Via G. Sansone 1, 50019 Sesto Fiorentino, Firenze, Italy}
\emailAdd{fares@iip.ufrn.br}
\emailAdd{raoul.santachiara@gmail.com}
\emailAdd{viti.jacopo@gmail.com}

\abstract{We study conformal twist field four-point functions on a $\mathbb Z_N$ orbifold. We examine in detail  the case $N=3$ and analyze theories obtained by replicated $N$-times a minimal model with central charge $c<1$. A fastly convergent expansion of the twist field correlation function in terms of sphere conformal blocks with central charge $Nc$ is obtained by exploiting covering map techniques. We discuss extensive applications of the formalism to the entanglement of two disjoint intervals in CFT, in particular we propose a conformal block expansion for the partially transposed reduced density matrix. Finally, we  refine the bounds on the structure constants of unitary CFTs determined previously by the genus two modular bootstrap.}

\begin{document} 
\maketitle
\flushbottom

\section{Introduction}
\label{sec:intro}
The conformal bootstrap approach in two-dimensions~\cite{BPZ} exploits the infinite-dimensional Virasoro algebra together with the associativity of the Operator Product Expansion (OPE) in the attempt to classify all possible Conformal Field Theories (CFTs). In short~\cite{Ribault}, the associativity of the conformal algebra leads to functional relations between its special functions dubbed conformal blocks, which can be solved, under certain assumptions on the spectrum of the theory, either analytically or numerically. 

In the last thirty years, the bootstrap technique allowed calculating exactly or with great precision correlation functions of many two-dimensional critical statistical models \cite{BPZ}. It also led to the  solution of two-dimensional quantum gravity, the Liouville theory~\cite{ZZ}. Despite these successes, there are still critical phenomena in two dimensions for which a complete bootstrap solution is lacking. We can mention for instance geometrical phase transitions such as percolation~\cite{DV}, recently analyzed in deep in~\cite{Picco, He}, or disordered systems~\cite{Ludwig, CardyBook}. In particular, the latter are described by  coupling  copies of the original pure CFT to a relevant field, and then flow to a new fixed point whose properties are largely unknown, see ~\cite{DotsenkoJacobsen, Komargodski, DelfinoRev}.

Besides their applications to disordered systems, CFTs on a replicated space-time geometry are familiar also in the context of quantum information. Partition functions of CFTs on Riemann surfaces with $\mathbb Z_N$ symmetry can be interpreted as powers of reduced density matrices for subsystems embedded into an extended quantum state, either pure or mixed~\cite{Holzhey, CalCardy04}. From the reduced density matrix, one extracts entanglement measures such as R\'enyi entropies~\cite{Caraglio, Calabrese09, Calabrese11} or the logarithmic negativity~\cite{Vidal, CalabreseNeg, CalabreseNeg12}.  
Finally, the study of  CFTs with cyclic or permutational symmetry has provided an explicit verification of the AdS/CFT correspondence~\cite{Gaberdiel}. A possible approach to CFTs defined on a $\mathbb Z_N$ (or $S_N$) symmetric space-time exploits their mapping~\cite{Lunin} to the orbifold theory ${\rm{CFT}}^{\otimes N}/\mathbb Z_N$. Instead of considering a single theory on a Riemann surface with $\mathbb Z_N$ symmetry and central charge $c$, one studies the $N$-fold tensor product ${\rm{CFT}}^{\otimes N}/\mathbb Z_{N}$ with central charge $Nc$ on the  sphere. The multivaluedness of the correlators, when their argument encircles a ramification point of the Riemann surface, is implemented by local fields called twist fields~\cite{Dixon, Knizhnik}. 
Higher genus partition functions are then mapped to multipoint correlators of twist fields.

In this paper, we will analyze the CFT partition function on a one-parameter ($0<x<1$) family of genus $N-1$ Riemann surfaces, specified  in Sec.~\ref{sec:surface} and already considered in~\cite{Headrick}. Due to the $\mathbb Z_N$ symmetry of the space-time, the CFT partition function can be expressed in terms of the  four-point correlator of the $\mathbb Z_N$ twist field on the sphere. Modular invariance of the partition function is implemented by the  transformation  $x\mapsto 1-x$  and reduces to the crossing symmetry of the twist field four-point function, see also~\cite{ZamolodchikovAT}. For $N=2$ and $N=3$, the study of the crossing symmetry equations restricts respectively  the operator content and  the OPE coefficients  of the original conformal theory~\cite{Cardy,CardyMod, Collier, Keller}, with central charge $c$,  on the sphere.

Already  for $\mathbb Z_3$-symmetric Riemann surfaces with genus two, the  determination of crossing symmetric twist field correlation functions is challenging.
For instance, they are not known explicitly even for the simplest CFTs such as the minimal models~\cite{BPZ} with $c<1$.  On the one hand,  if the seed CFT has a finite number of OPE channels, i.e. is rational, the twist field correlators expand over a finite number of special functions dubbed orbifold conformal blocks. On the other hand, differently from  what happens on the sphere, the latter cannot be calculated efficiently through a recursion relation~ \cite{Cho,Collier}. In practice, one  sets up a combinatorial expansion  of the orbifold conformal blocks, which produces a slowly convergent power series about $x=0$.  By truncating the power series in $x$, the crossing symmetry properties of the twist field correlator, and the modular invariance of the higher genus partition function, are poorly reproduced. Here, we revisit the problem by focussing on the minimal models. We propose a  systematic expansion of the $\mathbb Z_{3}$-orbifold conformal blocks that allows building crossing symmetric twist field four-point functions with significantly better accuracy than  previous attempts~\cite{Rajabpour, Ruggiero}. An alternative approach which involves the formulation of differential equations satisfied by the partition functions has been put forward in~\cite{Dupic}. Applications to entanglement measures and the modular bootstrap will be also investigated.

The outline of the paper is as follows. In Sec.~\ref{sec:surface}, we introduce our notations for a CFT on a Riemann surface with $\mathbb Z_N$ symmetry and briefly review the orbifold construction. The orbifold conformal blocks at $N=2,3$ and their small-$x$ expansion, obtained through the covering map~\cite{Lunin} procedure, are discussed in Sec.~\ref{sec:conf_blocks}. It will be further presented  a regularization prescription for the singularities in the power series that are produced, for $c<1$, by the states with zero norm.  
The contribution of the $\text{CFT}^{\otimes N}$ descendants to the small-$x$ expansion reorganizes~\cite{Headrick} into  conformal blocks of primary fields with respect to a Virasoro algebra with central charge $Nc$. Once this decomposition is found, we show in Sec.~\ref{sec:sphere_conf_blocks}, how the convergence of the orbifold conformal blocks can be improved~\cite{Rajabpour} by using the elliptic recursion formula~\cite{Zamolodchikov}. Refs.~\cite{Rajabpour, Ruggiero} wrote down analogous expansions in terms of Virasoro conformal blocks with central charge $Nc$ but they only took into account the orbifold primary contributions.

We first apply the formalism to  entanglement measures in tripartite systems at zero temperature.  In particular, we calculate in Sec.~\ref{sec:applications} power series representations of the trace of the third power of  the reduced density matrix and its  partial transpose for two disjoint intervals. The two traces are distinguished field theoretically by the presence of fields with non-zero conformal spins. Prior calculations of the partial transpose of the reduced density matrix, which enters in the so-called logarithmic negativity, were performed for free fermions~\cite{CalabreseNeg, Tagliacozzo, Alba, Eisler, Coser, Coser2, Coser3, Shapourian, Grava} or in the large central charge limit~\cite{Policastro, Ryu, Kusuki}. Our approach reproduces the free fermion results~\cite{Tagliacozzo}. Finally, in Sec.~\ref{sec:applications}, we show that for $N=3$, our expansion of the orbifold conformal blocks can incrementally improve  the OPE fusion coefficients bounds found in \cite{Collier}. Moreover, we demonstrate that, within our scheme, the numerical bootstrap approach to the genus two twist field correlation functions converges. These new results and observations can pave the way to the solution of a long-standing problem: the determination via a bootstrap approach of the mutual information and logarithmic negativities in interacting CFTs. 
\section{CFT Partition Functions on $\mathbb Z_N$-symmetric Riemann Surfaces}
\label{sec:surface}

Let us consider a CFT denoted by $\mathcal{C}$, with central charge 
$c$, defined on a Riemann surface $\Sigma_g(x)$  of genus $g$. In the following, 
we will refer to $\mathcal{C}$ as the seed theory.  We will restrict to the 
family of  Riemann surfaces $\Sigma_g(x)$ of genus $g=N-1$ described 
by the  complex algebraic curve
\begin{equation}\label{riemann_surf}
 w^N=\frac{z(z-1)}{z-x},
\end{equation}
which has branch points of order $N$ at $z_{\text{b}}=0$, $x$, $1$, and $\infty$. Although Eq.~\eqref{riemann_surf} can be extended to complex values of $x$, in the following, we will always assume $x$ to be a real variable, $x=\bar{x}$,  with $0<x<1$. Eq.~\eqref{riemann_surf} can be interpreted~\cite{Dubrovin} as a $N$-sheeted cover of the compactified complex plane (Riemann sphere) $\mathbb C\cup \{\infty\}$ with coordinate $z$. Furthermore, the surfaces $\Sigma_g(x)$ posses a $\mathbb{Z}_N$ symmetry since 
Eq.~\eqref{riemann_surf} is invariant under the transformation 
$w\mapsto e^{\frac{2\pi i k}{N}}w$, with $k=0,1,\dots,N-1$. 
This transformation amounts to a cyclic permutation of the $N$ sheets of the surface, where each sheet is labelled by the choice of the branch of the $N$-th root in Eq.~\eqref{riemann_surf}.

The CFT partition function  $\mathcal{Z}_{g}(x)$ on $\Sigma_{g}(x)$    depends on the choice of the metric  within the same conformal class.  We choose then a flat metric. The partition function  with a flat metric everywhere on the surface can be derived from the orbifold  $\mathcal{C}^{\otimes N}/\mathbb{Z}_N$~\cite{Lunin}. In this theory, there exist $\mathbb{Z}_N$ twist and anti-twist fields $\sigma_N$ and $\tilde{\sigma}_N$, which are  spinless primary fields of conformal dimension~\cite{Dixon, Knizhnik}
\begin{equation}
 h_{\sigma_N}=\bar{h}_{\sigma_N}=\frac{c}{24}\left(N-\frac{1}{N}\right).
\end{equation}
When inserted on the complex plane at the branch points of the algebraic curve in Eq.~\eqref{riemann_surf}, they implement the multivaluedness of correlation functions under the analytic continuation $(z-z_\text{b})\mapsto(z-z_\text{b})e^{2\pi i}$. One then finds that
\begin{equation}
\label{partition_twist_correl}
\mathcal{Z}_{g}(x)=e^{cS_{\text{anom.}}(x)}\langle \sigma_N (\infty) \tilde{\sigma}_N(1)\sigma_N(x,\bar{x}) \tilde{\sigma}_N(0)\rangle,
 \end{equation}
where the prefactor $e^{cS_{\text{anom.}}(x)}$ in Eq.~\eqref{partition_twist_correl} is the Weyl anomaly which can be explicitly calculated~\cite{Lunin}.
It takes into account that in the orbifold approach the metric employed to determine the partition function on $\Sigma_g(x)$ is a flat metric on each sheet of the surface but with conical singularities at the location of the twist fields. Consider, for instance, the case $N=2$, for which, under the Abel-Jacobi map~\cite{Miranda}, $\Sigma_{1}(x)$ is conformally equivalent to a flat torus of modulus 
\begin{equation}\label{tau}
 \tau(x)=i\frac{K(1-x)}{K(x)}, 
\end{equation}
where $K(x)$ is the complete elliptic integral of first 
kind~\cite{Whittaker}. By evaluating the conformal anomaly in Eq.~\eqref{partition_twist_correl}, one has~\cite{Lunin}
\begin{equation}\label{partition_torus_twist}
 \mathcal{Z}_1(x)
 =|2^8 x(1-x)|^{c/12} \langle \sigma_2 (\infty)\tilde{\sigma}_2(1)\sigma_2(x, \bar{x})\tilde{\sigma}_2(0)\rangle.
\end{equation}
The partition function $\mathcal{Z}_{g}(x)$ is invariant  
under modular transformations~\cite{CardyMod, Cappelli, Cappelli2}.  For the class of surfaces $\Sigma_{g}(x)$, the moduli space is one-dimensional and modular invariance implies the crossing
symmetry of the twist field four-point correlation function~\cite{Cardy}
\begin{equation}\label{cross_symmetry}
 \langle \sigma_N(\infty)\tilde{\sigma}_N(1)\sigma_N(1-x, 1-\bar{x})\tilde{\sigma}_N(0)\rangle=
 \langle \sigma_N(\infty)\tilde{\sigma}_N(1)\sigma_N(x, \bar{x})\tilde{\sigma}_N(0)\rangle. 
\end{equation}
For example, if we consider a torus with modulus $\tau$ that of Eq.~\eqref{tau},
then the modular transformation $\tau\mapsto-1/\tau$ implies
$x\mapsto 1-x$. An analogous observation holds for $\Sigma_{2}(x)$, as discussed in~\cite{Cardy}. Eq.~\eqref{cross_symmetry} can be actually extended analytically to complex values of $x$, see for instance Sec.~\ref{sec:bound_str_cts}.

\section{Orbifold Conformal Blocks}\label{sec:conf_blocks}
The twist field four-point correlator in Eq.~\eqref{partition_twist_correl} as a function of $x\in(0,1)$ can be analytically continued to $z\in\mathbb C$. In the orbifold $\mathcal{C}^{\otimes N}/\mathbb Z_N$, it admits the following decomposition
\begin{equation}\label{conformal_block_decomposition_0}
 \langle \sigma_N (\infty)  \tilde{\sigma}_N(1) \sigma_N (z, \bar{z})\tilde{\sigma}_N(0)\rangle=
 \sum_{\boldsymbol{h},\boldsymbol{\bar{h}}} D_{\boldsymbol{h}, \boldsymbol{\bar{h}}}
 \mathcal{G}_{c, \boldsymbol{h}}^{(N)}(z)\mathcal{G}_{c,\boldsymbol{\bar{h}}}^{(N)}(\bar{z}),
\end{equation}
where $\boldsymbol{h}\equiv\{h_1, \dots, h_N\}$ and $h_j$ is the conformal dimension of  
a primary field of the seed theory. The functions $\mathcal{G}_{c, \boldsymbol{h}}^{(N)}(z)$, 
defined below, will be termed orbifold conformal blocks. They are normalized such that for small $|z|$
\begin{equation}
\label{G_asy}
\mathcal{G}_{c, \boldsymbol{h}}^{(N)}(z)=z^{|\boldsymbol{h}| -2h_{\sigma_N}}\left[1+ O(z)\right],
\end{equation}
where $|\boldsymbol{h}|=\sum_j h_j$. In this section and Sec.~\ref{sec:sphere_conf_blocks}, we will show how to extract systematically, by means of the orbifold conformal algebra, higher order terms in the expansion about $z=0$ in Eq.~\eqref{G_asy}. The case $N=3$ has been discussed in~\cite{Collier} from which some notations are borrowed. The structure constants  $D_{\boldsymbol{h}, \boldsymbol{\bar{h}}}$ in Eq.~\eqref{conformal_block_decomposition_0} are not algebraically determined and instead  characterize the specific bootstrap solution under consideration.

We focus first on the holomorphic sector of the seed CFT. We denote by $\phi_{h}(z)$  the holomorphic primary field with conformal dimension $h$  and by $\phi_{h}^{M}(z)$ one of its descendants. The descendants are labelled by the partition  $M\equiv\{m_1,\dots,m_q\}$, $1\leq m_1\leq m_2\cdots\leq m_q$, of the positive integer  $|M|=\sum_{j} m_j$. In terms of the  Virasoro generators $L_{-m}(z)$, defined in Eq.~\eqref{vir_ln}, the holomorphic field $\phi_{h}^{M}(z)$ is then
\begin{equation}
\label{desc}
 \phi_{h}^{M}(z)=L_{-M}(z)\;\phi_h(z),~~{\rm{where}}~~ L_{-M}(z)=L_{-m_1}(z)\dots L_{-m_q}(z).
 \end{equation}
The field  $\phi_{h}^{M}(z)$ has conformal dimension $h+|M|$. 

We will employ the field-state correspondence $|\phi_{h}\rangle\equiv \lim_{z\to 0}\phi_h(z)|0\rangle$, with $|0\rangle$ the vacuum in $\mathcal{C}$, and the Virasoro scalar product \cite{DiFrancesco}. 
The latter can be defined  by constructing the dual Hilbert space  through the identification  $\langle \phi_{h} |\equiv\lim_{z\rightarrow\infty}z^{2h}\langle 0|\phi_h(z)$, where $\langle 0|$ is the dual of the vacuum state. Furthermore, we denote by  $G^{h}_{M_1,M_2}$ the matrix of scalar products
\begin{equation}
G^{h}_{M_1,M_2} = \langle \phi_{h}^{M_1}|\phi_{h}^{M_2}\rangle. 
\end{equation}
If not  stated otherwise, in what follows we consider irreducible Verma module representations~\cite{Ribault}, i.e. there are not descendant states with vanishing norm, referred here as null vectors. The role of null vectors will be analyzed in more detail in Sec.~\ref{null_vec1}.  The fields that enter the correlation functions are specified by gluing the holomorphic and anti-holomorphic sectors of $\mathcal{C}$ and are tensor products of the form
\begin{equation}
\label{fields}
\phi^{M}_{h}(z)\phi^{\bar{M}}_{\bar{h}}(\bar{z}).
\end{equation}
In all applications, we will actually restrict ourselves to the case of diagonal seed CFTs, i.e. theories that have a diagonal partition function on the torus~\cite{Cappelli}. In this case, in Eq.~\eqref{fields}, only couplings between holomorphic and antiholomorphic fields with the same labels $h=\bar{h}$ are possible. However, in this section we prefer to maintain the discussion more general and include the possibility of non-diagonal couplings.
In the tensor product $\mathcal{C}^{\otimes N}$, an holomorphic primary $\boldsymbol{\phi_{h}}$, labelled by a set $\boldsymbol{h}$ of conformal dimensions, is the tensor product of holomorphic primary fields of the seed theory
\begin{equation}
\boldsymbol{\phi}_{\boldsymbol{h}}(z)=\phi_{h_1}(z)\otimes \cdots\otimes \phi_{h_N}(z),
\end{equation}
and has conformal dimension $|\boldsymbol{h}|$. If $\boldsymbol{M}\equiv\{M_1,\dots,M_N\}$ stands for a collection of $N$ partitions of the positive integers $|M_1|,\dots,|M_N|$, then the descendants of $\boldsymbol{\phi}_{\boldsymbol{h}}$ will be indicated by $\boldsymbol{\phi}^{\boldsymbol{M}}_{\boldsymbol{h}}(z)$,
 \begin{equation}
 \label{descor}
 \boldsymbol{\phi}_{\boldsymbol{h}}^{\boldsymbol{M}}(z)=\phi_{h_1}^{M_1}(z)\otimes \dots \otimes \phi_{h_N}^{M_N}(z).
\end{equation}
To complete the fields in Eq.~\eqref{descor} to a basis for a representation of the tensor product of the Virasoro algebra, we shall allow $M_{j}$ to be the empty set, with the convention that $\phi_{h_j}^{\varnothing}\equiv\phi_{h_j}$.
The corresponding scalar product matrix $\boldsymbol{G}^{\boldsymbol{h}}_{\boldsymbol{M}\boldsymbol{N}}$, of size $\prod_{j=1}^{N}|M_j|^2\times \prod_{j=1}^{N}|N_j|^2$, is defined from the scalar product in $\mathcal{C}$ as
\begin{equation}
\label{scal}
 \boldsymbol{G}^{\boldsymbol{h}}_{\boldsymbol{M}\boldsymbol{N}}=\langle \boldsymbol{\phi}_{\boldsymbol{h}}^{\boldsymbol{M}} | \boldsymbol{\phi}_{\boldsymbol{h}}^{\boldsymbol{N}}\rangle=\prod_{j=1}^N G^{h_j}_{M_j,N_j}.
\end{equation}
The construction of the anti-holomorphic sector is, under the replacement $z\to \bar{z}$, the same as the one presented above and  the  fields in $\mathcal{C}^{\otimes N}$ are then the tensor products
$\boldsymbol{\phi}^{\boldsymbol{M}}_{\boldsymbol{h}}(z)\boldsymbol{\phi}^{\boldsymbol{\bar{M}}}_{\boldsymbol{\bar{h}}}(\bar{z})$.
The expansion about $z=0$ of $\mathcal{G}_{c, \boldsymbol{h}}^{(N)}(z)$ can be determined by inserting the resolution of the identity in the representation $\boldsymbol{h}$. By using the basis in Eq.~\eqref{descor}, i.e. including among the elements of $\boldsymbol{M}$ and $\boldsymbol{N}$ also  the empty set, it follows that
\begin{equation}\label{conformal_block_decomposition}
 \mathcal{G}_{c,\boldsymbol{h}}^{(N)}(z)=
 z^{|\boldsymbol{h}|-2h_{\sigma_N}}\sum_{\boldsymbol{M}, \boldsymbol{N}} z^{|\boldsymbol{M}|}\;\tilde{\rho}^{\boldsymbol{h}}_{\boldsymbol{M}}\; [\boldsymbol{G}^{\boldsymbol{h}}_{\boldsymbol{M},\boldsymbol{N}}]^{-1}\;\rho^{\boldsymbol{h}}_{\boldsymbol{N}},
\end{equation}
where $\rho_{\boldsymbol{N}}^{\boldsymbol{h}}$ and $\tilde{\rho}^{\boldsymbol{h}}_{\boldsymbol{M}}$ are matrix elements between descendant fields. In terms of the orbifold structure constants,
\begin{equation}
\label{strucconst}
C_{\boldsymbol{h},\boldsymbol{\bar{h}}}\equiv\langle \boldsymbol{\phi}_{\boldsymbol{h}}\boldsymbol{\phi}_{\boldsymbol{\bar{h}}}| \sigma_{N}(1)|\tilde{\sigma}_N\rangle, \quad 
\tilde{C}_{\boldsymbol{h}, \boldsymbol{\bar{h}}}\equiv\langle \boldsymbol{\phi}_{\boldsymbol{h}}\boldsymbol{\phi}_{\boldsymbol{\bar{h}}}|\tilde{\sigma}_N(1)|\sigma_N\rangle,
\end{equation}
one has
 \begin{equation}
 \label{Gammas}
\rho^{\boldsymbol{h}}_{\boldsymbol{M}} = \frac{\langle \boldsymbol{\phi}^{\boldsymbol{M}}_{\boldsymbol{h}} \boldsymbol{\phi}_{\boldsymbol{\bar{h}}}| \sigma_{N}(1)|\tilde{\sigma}_{N}\rangle}{C_{\boldsymbol{h},\boldsymbol{\bar{h}}}},\quad  
\tilde{\rho}^{\boldsymbol{h}}_{\boldsymbol{M}} = \frac{\langle\boldsymbol{\phi}^{\boldsymbol{M}}_{\boldsymbol{h}} \boldsymbol{\phi}_{\boldsymbol{\bar{h}}}|\tilde{\sigma}_{N}(1)|\sigma_N\rangle}{\tilde{C}_{\boldsymbol{h},\boldsymbol{\bar{h}}}} .
 \end{equation}

 The matrix elements  $\rho^{\boldsymbol{h}}_{\boldsymbol{M}}$ and $\tilde{\rho}^{\boldsymbol{h}}_{\boldsymbol{M}}$ are entirely fixed by the holomorphic part of the orbifold conformal algebra. Even if it is not manifest in their expressions in Eq.~\eqref{Gammas}, they are complex rational functions of the dimensions $h_j$ and  the central charge $c$. We will show how to compute them in the next section. Due to the symmetry properties of the cyclic twist and anti-twist fields, $\tilde{\rho}_{\boldsymbol{M}}^{\boldsymbol{h}}$ is the complex conjugate of $\rho_{\boldsymbol{M}}^{\boldsymbol{h}}$.
 
 The structure constants in Eq.~\eqref{strucconst}, on the other hand, encode the way the holomorphic and anti-holomorphic sector are glued to build the twist correlation function of the model under consideration. Plugging Eq.~\eqref{conformal_block_decomposition} into Eq.~\eqref{conformal_block_decomposition_0}, one concludes  that 
 \begin{equation}\label{DC}
 D_{\boldsymbol{h},\boldsymbol{\bar{h}}}= C_{\boldsymbol{h},\boldsymbol{\bar{h}}}\tilde{C}_{\boldsymbol{h}, \boldsymbol{\bar{h}}}.
 \end{equation}
\subsection{The Computation of the Orbifold Three-point Functions}
The computation of the twist field four-point function boils down to determine the 
orbifold three-point functions, see Eq.~\eqref{Gammas},
\begin{equation}\label{orbifold_three-point}
 \rho_{\boldsymbol{M}}^{\boldsymbol{h}} \;C_{\boldsymbol{h},\boldsymbol{\bar{h}}} = \langle \boldsymbol{\phi}^{\boldsymbol{M}}_{\boldsymbol{h}} \boldsymbol{\phi}_{\boldsymbol{\bar{h}}}|\sigma_{N}(1)|\tilde{\sigma}_{N}\rangle.
\end{equation}
These quantities can be calculated by considering a $N$-to-one conformal map
$t\mapsto z(t)$ with branch points at $z_{\text{b}}=\{0, 1\}$ such that, 
near these points, it behaves as $z-z_{\text{b}}\sim (t-t_{\text{b}})^N$.
That is, the $t$-surface must be a $N$-sheeted cover with genus zero, a Riemann sphere, of the complex plane with a branch cut along 
$z\in(0,1)$.  Moreover, the point $z=\infty$, where the holomorphic field $\boldsymbol{\phi}^{\boldsymbol{M}}_{\boldsymbol{h}}$
in Eq.~\eqref{orbifold_three-point} is inserted, must be mapped to $N$ different points $t_\infty=\{t_1, \dots, t_N\}$ in 
the covering space. Let us denote by $\hat{\phi}_{h_j}^{M_j}(t_j)$ the images
of the field $\phi_{h_j}^{M_j}(\infty)$ under the covering map $t\mapsto z(t)$, that is
\begin{equation}
\label{Jac}
\hat{\phi}_{h_j}^{M_j}(t_j) = \left.\left(\frac{d z(t)}{d t}
 \right)^{-h_j}\right|_{t=t_j} \mathcal{L}_{-M_j}(t_j)\;\phi_{h_j} (t_j),
\end{equation}
where $\mathcal{L}_{-M_j}(t_j)$ is the pullback of the Virasoro operator
$L_{-M_j}(z=\infty)$, see Eq.~\eqref{Virasoro_transf}. The pullback is a linear combination of Virasoro generators
acting at the point $t_j$ in the $t$-plane, as we discuss in detail in Appendix~\ref{app_ward}.
The holomorphic part of the three-point function of the $\mathcal{C}^{\otimes N}/\mathbb{Z}_N$ orbifold in Eq.~\eqref{orbifold_three-point}
is then equal to the $N$-point function of the seed theory $\mathcal{C}$ on a sphere
\begin{equation}\label{N-point}
\langle \boldsymbol{\phi}^{\boldsymbol{M}}_{\boldsymbol{h}}(\infty) \sigma_{N}(1)\tilde{\sigma}_{N}(0)\rangle=\langle \hat{\phi}^{M_1}_{h_1}(t_1)\cdots\hat{\phi}_{h_N}^{M_N}(t_N)\rangle.
\end{equation}

\subsection{Case $N=2$: The Torus}\label{sec:genus_one}

We now illustrate the method discussed in the previous section when $N=2$. 
In this case, the orbifold three-point function of Eq.~\eqref{orbifold_three-point} 
can be calculated by considering the two-to-one map
\begin{equation}\label{g_one_covering_map}
 z(t)=\frac{(t+1)^2}{4t},
\end{equation}
which transforms $t_{\infty}=\{0,\infty\}$ into $z=\infty$ and has branch points of 
order two at $z_{\text{b}}=\{0, 1\}$. By applying Eq.~\eqref{Jac} and Eq.~\eqref{N-point}, the three-point functions in Eq.~\eqref{orbifold_three-point} reduce to scalar products.  In particular, one has,  see also Eq.~\eqref{strucconst}, 
\begin{align}
\label{C_torus}
 C_{\boldsymbol{h},\boldsymbol{\bar{h}}}=
 \langle \hat{\phi}_{h_1}\hat{\phi}_{\bar{h}_1}|\hat{\phi}_{h_2}\hat{\phi}_{\bar{h}_2}\rangle
 =\delta_{h_1, h_2}\delta_{\bar{h}_1,\bar{h}_2}2^{-4(h_1+\bar{h}_1)},
\end{align}
while $\rho^{h_1,h_2}$ is the symmetric $|M_1|\times|M_1|$ matrix (descendant fields at different levels are orthogonal)
\begin{equation}
\label{rho}
 \rho^{h_1,h_2}_{M_1,M_2}=\langle \mathcal{L}_{-M_1}\phi_{h_1}|\mathcal{L}_{-M_2}\phi_{h_2}\rangle\delta_{h_1,h_2}.
\end{equation}
The explicit expression of the pullback $\mathcal{L}_{-n}$ under the map of Eq.~\eqref{g_one_covering_map}
is written in Eq.~\eqref{vir_g1_1} of Appendix~\ref{app_ward}. The matrix $\rho^{h_1,h_2}$ in Eq.~\eqref{rho} can be easily calculated by exploiting the Virasoro algebra commutation relations. 
Moreover, in the case $N=2$, the twist and anti-twist fields are identified ($\sigma_2=\tilde{\sigma}_2$) and, from Eqs.~\eqref{strucconst} and \eqref{Gammas}, we conclude that $\tilde{C}_{\boldsymbol{h}, \boldsymbol{\bar{h}}}=C_{\boldsymbol{h}, \boldsymbol{\bar{h}}}$ and $\tilde{\rho}_{M_1, M_2}^{h_1, h_2}=\rho_{M_1, M_2}^{h_1, h_2}$.

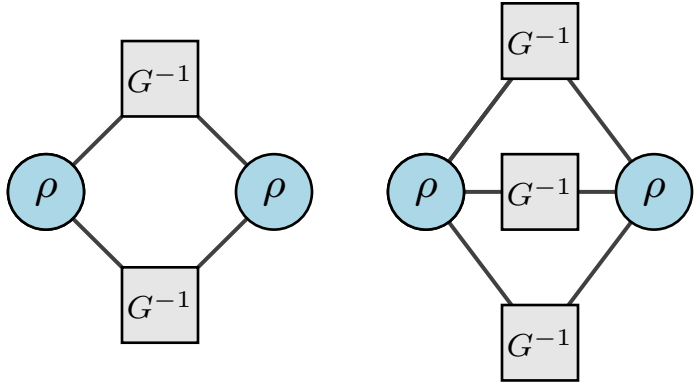
\begin{figure}[t]
\centering
 \begin{tikzpicture}
  \Vertex[size=1, label=$\rho$, fontscale=2]{A};
  \Vertex[x=1.5, y=1.5, size=1, shape=rectangle, color=gray!20, label=$G^{-1}$, fontscale=1.5]{B};
  \Vertex[size=1, label=$\rho$, fontscale=2]{A};
  \Vertex[x=3, y=0, size=1, label=$\rho$, fontscale=2]{C};
  \Vertex[x=1.5, y=-1.5, size=1, shape=rectangle, color=gray!20, label=$G^{-1}$, fontscale=1.5]{D};
  \Edge(A)(B);
   \Edge(B)(C);
    \Edge(C)(D);
     \Edge(D)(A);
     \begin{scope}[xshift= 5cm]
      \Vertex[size=1, label=$\rho$, fontscale=2]{A};
  \Vertex[x=1.5, y=2, size=1, shape=rectangle, color=gray!20, label=$G^{-1}$, fontscale=1.5]{B};
  \Vertex[size=1, label=$\rho$, fontscale=2]{A};
  \Vertex[x=3, y=0, size=1, label=$\rho$, fontscale=2]{C};
  \Vertex[x=1.5, y=-2, size=1, shape=rectangle, color=gray!20, label=$G^{-1}$, fontscale=1.5]{D};
  \Vertex[x=1.5, y=0, size=1, shape=rectangle, color=gray!20, label=$G^{-1}$, fontscale=1.5]{E};
  \Edge(A)(B);
  \Edge(A)(B);
   \Edge(B)(C);
    \Edge(C)(D);
     \Edge(D)(A);
     \Edge (A)(E);
     \Edge (E)(C);
     \end{scope}

 \end{tikzpicture}
\caption{On the left: Matrix product representation of the genus one conformal block in Eq.~\eqref{genus_one_conf_block}. On the right: Matrix product representation of the genus two conformal block in Eq.~\eqref{genus_two_conf_block}.}
\label{fig-g1}
\end{figure}

Finally, by substituting Eq.\eqref{rho} in Eq.\eqref{conformal_block_decomposition}, we  arrive at a power series expansion of the $N=2$ orbifold conformal block
\begin{equation}\label{genus_one_conf_block}
 \mathcal{G}_{c, \{h_1,h_2\}}^{(2)}(z)=\delta_{h_1,h_2}z^{h_1+h_2-\frac{c}{8}}\sum_{M_1, M_2}\sum_{N_1, N_2}
 z^{|M_1|+|M_2|}~\rho^{h_1,h_2}_{M_1,M_2}~\prod_{j=1}^{2}[G^{h_j}_{M_j,N_j}]^{-1}
~\rho^{h_1,h_2}_{N_1,N_2}.
\end{equation}
By recalling that the matrices $G^{h_j}$ and $\rho^{h_1,h_2}$ are symmetric, the coefficients of the combinatorial 
expansion in Eq.~\eqref{genus_one_conf_block} can be organized as matrix products, presented in the diagram 
of Fig.~\ref{fig-g1}. This observation  helps with the organization of the bookkeeping of the states.

Eq.~\eqref{genus_one_conf_block} shows that the number of $N=2$ orbifold conformal blocks is the same as the number
of conformal families of the seed theory $\mathcal{C}$ . This conclusion is of course consistent with
the well known construction of the CFT partition function on a flat torus as a sesquilinear form of the
irreducible Virasoro characters $\chi_{c,h}(\tau(x))$~\cite{Cappelli, Cappelli2},
\begin{equation}
\label{pf}
\mathcal{Z}_1(x) = \sum_{h,\bar{h}}n_{h,\bar{h}}~\chi_{c,h}(\tau(x)) \chi_{c,\bar{h}}(\bar{\tau}(x)).
\end{equation}
We can relate the conformal block in Eq.~\eqref{genus_one_conf_block} and the Virasoro character by considering theories with a diagonal partition function on the torus. In this case, the multiplicities in Eq.~\eqref{pf} are $n_{h,\bar{h}}=\delta_{h,\bar{h}}$ and only couplings between holomorphic and anti-holomorphic fields belonging to the same Virasoro algebra representation are allowed. Eq.~\eqref{C_torus} must be now supplemented by the constraint $\boldsymbol{h}=\boldsymbol{\bar{h}}$ and we then derive from Eq.~\eqref{DC}
\begin{equation}\label{D_2}
 D_{\boldsymbol{h},\boldsymbol{\bar{h}}} =\delta_{h_1,h_2}\;\delta_{\bar{h}_1,\bar{h}_2}\delta_{\boldsymbol{h},\boldsymbol{\bar{h}}} \;2^{-16 h_1}.
\end{equation}
 By substituting Eq.~\eqref{D_2} into Eq.~\eqref{conformal_block_decomposition_0} and then comparing Eq.~\eqref{partition_torus_twist} with Eq.~\eqref{pf}, one eventually identifies~\cite{Headrick}
\begin{equation}\label{character_conf_block}
\mathcal{G}_{c,\{h, h\}}^{(2)}(x)=2^{8h-c/3}[x(1-x)]^{-c/24}\chi_{c, h}(\tau(x)),
\end{equation}
which we will also use in Sec.~\ref{sec:null_vectors}.

\subsection{Case $N=3$: $\mathbb Z_3$-symmetric Riemann Surfaces with Genus Two}\label{sec:genus_two}

For $N=3$, the orbifold three-point function in Eq.~\eqref{orbifold_three-point} can be computed 
by introducing, for instance, the three-to-one map~\cite{Collier}
\begin{equation}\label{g_two_covering_map}
 z(t) =\frac{(t+\omega)^3}{3\omega(1-\omega)t(t-1)},\quad \omega=e^{\frac{2\pi i}{3}}.
\end{equation}
This transformation has branch points of order three at 
$z_{\text{b}}=\{0,1\}$ and maps the points $t_{\infty}=\{0, 1, \infty\}$ in the $t$-surface into $z=\infty$. The $t$-surface, which has again the topology of a sphere, is then a triple covering of the complex plane with a cut along $z\in(0,1)$. By recalling Eq.~\eqref{strucconst} and applying Eqs.~\eqref{Jac} and \eqref{N-point} one has
\begin{equation}
\label{C_3_0}
 C_{\boldsymbol{h}, \boldsymbol{\bar{h}}}=
 \langle \hat{\phi}_{h_1}\hat{\phi}_{\bar{h}_1}|\hat{\phi}_{h_2}(1)\hat{\phi}_{\bar{h}_2}(1)|\hat{\phi}_{h_3}\hat{\phi}_{\bar{h}_3}\rangle=[3\omega(1-\omega)]^{-h_1-h_2-h_3}[3\omega^2(1-\omega^2)]^{-\bar{h}_1-\bar{h}_2-\bar{h}_3}C_{\boldsymbol{h}, \boldsymbol{\bar{h}}}^{{\rm seed}}.
\end{equation}
where the structure constants $C_{\boldsymbol{h}, \boldsymbol{\bar{h}}}^{{\rm seed}}$ are calculated in the seed CFT, namely
\begin{equation}
 C_{\boldsymbol{h}, \boldsymbol{\bar{h}}}^{{\rm seed}}=\langle \phi_{h_1}\phi_{\bar{h}_1}|\phi_{h_2}(1)\phi_{\bar{h}_2}(1)|\phi_{h_3}\phi_{\bar{h}_3}\rangle.
\end{equation}
Analogously from Eq.~\eqref{Gammas}, it follows
\begin{equation}
\label{rho2}
 \rho^{h_1,h_2,h_3}_{M_1,M_2,M_3}=
 \frac{\langle \mathcal{L}_{-M_1}\phi_{h_1}|\mathcal{L}_{-M_2}(1)\phi_{h_2}(1)|\mathcal{L}_{-M_3}\phi_{h_3}\rangle}
 { \langle \phi_{h_1}|\phi_{h_2}(1)|\phi_{h_3}\rangle}.
\end{equation}
The pullback $\mathcal{L}_{-n}(t_{\infty})$ of the Virasoro descendant $L_{-n}(z=\infty)$ 
under the conformal map in Eq.~\eqref{g_two_covering_map} are obtained, as in the 
case $N=2$, by using Eq.~\eqref{Virasoro_transf}, see also Eq.~\eqref{vir_g2}.
The elements of Eq.~\eqref{rho2} can be calculated by implementing in a computer 
algebra system the Ward identities of Appendix~\ref{app_ward}. Finally, the structure constants $\tilde{C}_{\boldsymbol{h}, \boldsymbol{\bar{h}}}$ 
and matrix elements $\tilde{\rho}_{M_1, M_2, M_3}^{\;h_1, h_2, h_3}$, which follow by exchanging twist with anti-twist fields in Eqs.~\eqref{strucconst}  and \eqref{Gammas}, are determined by considering the complex conjugate 
of the map of Eq.~\eqref{g_two_covering_map}. One  finds
\begin{equation}\label{C_3_0_bar}
 \tilde{C}_{\boldsymbol{h}, \boldsymbol{\bar{h}}}=[3\omega^2(1-\omega^2)]^{-h_1-h_2-h_3}[3\omega(1-\omega)]^{-\bar{h}_1-\bar{h}_2-\bar{h}_3}C_{\boldsymbol{h}, \boldsymbol{\bar{h}}}^{{\rm seed}},
\end{equation}
and
\begin{equation}
\tilde{\rho}_{M_1, M_2, M_3}^{\; h_1, h_2, h_3}=[\rho_{M_1, M_2, M_3}^{h_1, h_2, h_3}]^*.
\end{equation}
The resulting expressions for the tensor $\rho_{M_1,M_2,M_3}^{h_1,h_2,h_3}$ in Eq.~\eqref{rho2} can be then eventually plugged into  Eq.~\eqref{conformal_block_decomposition} to get the expansion
\begin{equation}\label{genus_two_conf_block}
 \mathcal{G}_{c, \{h_1, h_2, h_3\}}^{(3)}(z)=
z^{\sum_{j=1}^3 h_j-\frac{2c}{9}}\sum_{\substack{\{M_j\}, \{N_j\}}}
 z^{\sum\limits_{j=1}^3 |M_j|}
  \rho^{h_1,h_2,h_3}_{M_1,M_2,M_3}~\prod_{j=1}^3 [G_{M_j,N_j}^{h_j}]^{-1}
~[\rho_{N_1,N_2,N_3}^{h_1,h_2,h_3}]^*,
\end{equation}
which is also illustrated pictorially in Fig.~\ref{fig-g1}. The conformal block $\mathcal{G}_{c, \{h_1, h_2, h_3\}}^{(3)}$ in Eq.~\eqref{genus_two_conf_block} is
manifestly symmetric under permutations of $h_1$, $h_2$ and $h_3$, consistently with the $\mathbb Z_3$ 
(in fact $S_3$) symmetry of the orbifold CFT. From Eq.~\eqref{C_3_0}, it follows that the 
$N=3$ orbifold conformal blocks are in one-to-one correspondence~\cite{Cardy, Collier} with the non-zero 
structure constants of $\mathcal{C}$. Indeed, if we insert Eqs.~\eqref{C_3_0} and \eqref{C_3_0_bar} 
into Eq.~\eqref{DC}, we conclude that, see also Eq.~\eqref{conformal_block_decomposition_0}, 
\begin{equation}\label{D_3}
 D_{\boldsymbol{h}, \boldsymbol{\bar{h}}}=
 27^{-|\boldsymbol{h}|-|\boldsymbol{\bar{h}}|}
 (C_{\boldsymbol{h}, \boldsymbol{\bar{h}}}^{{\rm seed}})^2.
\end{equation}

\subsection{Minimal Model Orbifold Conformal Blocks}\label{sec:null_vectors}
\label{null_vec1}
 We consider now a seed theory which is built upon irreducible representations of the Virasoro algebra, labelled by $c$ and $h$, constructed by removing all the null vectors~\cite{Ribault}. 
A prominent example of such CFTs are the minimal models~\cite{BPZ}  which have central charge
\begin{equation}\label{central_charge_min_mod}
c=c_{p,q}= 1 -6\frac{ (p-q)^2}{p q},
\end{equation}
being $p$ and $q$ positive coprime integers. The minimal models are rational theories, that is they contain a finite number of primaries $\phi_{h^{p,q}_{r,s}}$  with conformal dimensions
 \begin{equation}\label{conf_dim_min_mod}
 h^{p,q}_{r,s}=\frac{(pr-qs)^2-(p-q)^2}{4 p q}, \quad 1\leq r\leq q-1,  1\leq s\leq p-1,
 \end{equation}
 whose OPE algebra closes. We will denote by  $\boldsymbol{h}_{\text{deg}}\equiv\{h_{r_j, s_j}^{p, q}\}_{j=1}^N$ a set of $N$ conformal dimension as in Eq.~\eqref{conf_dim_min_mod}, the orbifold conformal block for a minimal model is then the function $\mathcal{G}_{c_{p,q}, \boldsymbol{h}_{\text{deg}}}^{(N)}(z)$.

Due to the chain of resonances $h_{r,s}^{p, q}=h_{q-r,p-s}^{p, q}=h_{q+r,p+s}^{p, q}=\cdots$, the Verma module labelled by $c_{p,q}$ and  $h^{p,q}_{r,s}$ has an infinite series of null vectors at the levels $r s, (q-r)(p-s),(q+r)(p+s), \dots$. As already emphasized, these states are absent from the spectrum of $\mathcal{C}$. Consequently, when calculating the  combinatorial expansion in Eq.~\eqref{conformal_block_decomposition} in a minimal model, one has to choose a basis of descendants for each representation $h_{r_j, s_j}^{p, q}$ which is free of null vectors. Even if there are closed expressions for the null vectors, the construction  of such a basis becomes cumbersome at higher levels, due to the resonances mentioned above. We refer to~\cite{Javerzat} where this particular issue is discussed in more detail. Here we will follow an alternative path, inspired by the AGT approach to the minimal models~\cite{Tanzini,Alkalaev,FodaAGT}, see also~\cite{SV}.

The analytic properties of $\mathcal{G}_{c, \boldsymbol{h}_{\text{deg}}}^{(N)}(z)$ for $N=2$ and $N=3$ as a function of $c$ can be  understood by examining Eqs.~\eqref{genus_one_conf_block}~and~\eqref{genus_two_conf_block} respectively. If there is a null vector at a certain level in the Verma module with conformal dimension $h_{r_j, s_j}^{p, q}$, the matrix elements of $\left[G^{h^{p,q}_{r_j,s_j}}\right]^{-1}$ have  a simple pole.  More precisely, they are order $O((c-c_{p, q})^{-1})$ since the null vector is orthogonal to all the states belonging to that level.  At the same time, each factor of $\rho$ produces a single zero, namely is order $O(c-c_{p,q})$. Indeed, for $N=2$, the elements of $\rho$, see  Eq.~\eqref{rho}, are scalar products and therefore vanish when evaluated with a null vector. For $N=3$, the vanishing  of the tensor $\rho^{h_{r_1,s_1},h_{r_2,s_2},h_{r_3,s_3}}$ is instead a consequence of the fusion rules of the minimal model, see Eq.~\eqref{rho2}.  For eliminating the null vector contribution from Eqs.~\eqref{genus_one_conf_block} and \eqref{genus_two_conf_block}, 
it is enough, if $\boldsymbol{h}\neq \boldsymbol{0}$ where $\boldsymbol{0}\equiv \{0,\dots, 0\}$, to regularize the central charge differently in the $\rho$ and $G$ factors. In particular, we take
\begin{equation}\label{regularization}
 \rho^{\boldsymbol{h}_{\text{deg}}}(c)\rightarrow\rho^{\boldsymbol{h}_{\text{deg}}}(c_{p,q}+\varepsilon^2),\quad  G^{h_{r_j, s_j}^{p, q}}(c)\rightarrow G^{h_{r_j, s_j}^{p, q}}(c_{p,q}+\varepsilon),
\end{equation}
with $\varepsilon>0$, in such a way that the null vectors contributions in Eq.~\eqref{conformal_block_decomposition} are now zero
in the limit $\varepsilon\to 0$. By recalling Eq.~\eqref{character_conf_block}, it is possible to check that at $N=2$, the power series about $z=0$ of the regularized
conformal block reproduces the analogous expansion of the character $\chi_{c_{p, q}, h_{r, s}^{p, q}}(\tau(z))$ in the minimal model~\cite{DiFrancesco}.

The case $\boldsymbol{h}=\boldsymbol{0}$ is  slightly different since the Verma module of the identity contains a null vector already at level one for any value of the central charge. Hence, we can follow 
a similar strategy but modifying the dependence of the matrices $G^{h}$ and $\rho^{\boldsymbol{h}}$
on the conformal dimensions rather than the central charge. We replace then 
\begin{equation}\label{regularization_identity}
 \rho^{\boldsymbol{h}=\boldsymbol{0}}(c)\rightarrow\rho^{\boldsymbol{h}=\{\delta^2,\dots, \delta^2\}}(c),~ G^{h=0}(c)\rightarrow G^{h=\delta}(c),
\end{equation}
with $\delta>0$ in Eq.~\eqref{conformal_block_decomposition}. The limit $\delta\to 0$ defines the regularized conformal block 
$\mathcal{G}_{c, \boldsymbol{0}}^{(N)}(z)$ for $N=2,3$. In particular, it provides, 
for $N=2$ upon using Eq.~\eqref{character_conf_block}, the  character of the irreducible 
representation of the identity. When $N=3$, note that, if the identity representation is present in only one of the replicas, 
for instance $h_1=0$, then, according to Eq.~\eqref{C_3_0},  we shall have the same conformal family in the 
other two replicas, i.e. $h_2=h_3$. In order to find the correct conformal block, in this case, 
one must  identify first $h_2=h_3$, while keeping $h_1\neq 0$, and then perform the limit $h_1\to 0$. 

Let us see how the regularization procedure in Eqs.~(\ref{regularization}, \ref{regularization_identity}) works in a particular example: the Ising CFT, which corresponds to $p=4$ and $q=3$, namely $c_{4, 3}=1/2$ in Eq.~\eqref{central_charge_min_mod}. For instance, for $h_{1, 2}=1/16$, 
the first few terms of the regularized expansion of Eq.~\eqref{genus_one_conf_block} about $z=0$ are 
\begin{equation}
 \mathcal{G}_{\frac{1}{2}, \{\frac{1}{16}, \frac{1}{16}\}}^{(2)}(z)=z^{1/16}
 \left[1+\frac{z}{16}+\frac{17z^2}{512}+\frac{187z^3}{8192}+\frac{9163z^4}{524288}+O(z^5)\right].
\end{equation}
The coefficient of the $O(z^4)$ term would be different if the regularization scheme in Eq.~\eqref{regularization} was not implemented.
Indeed the Verma module with conformal dimension $h=1/16$ at $c=1/2$ possesses a null vector at level two. If this vector appeared on both
replicas of the seed theory, i.e. in the sector $|M_1|=|M_2|=2$ of Eq.~\eqref{genus_one_conf_block}, it 
 would modify the coefficient of the  $O(z^4)$ term  in the conformal block. 
After implementing the regularization scheme in Eq.~\eqref{regularization}, its contribution is instead of order $O(\varepsilon^2)$
and drops in the limit $\varepsilon\to 0$.

As a second example, we analyze the regularized $N=3$ orbifold conformal block for $c=1/2$, $h_1=h_2=1/16$
and $h_3=0$; the result of the expansion in Eq.~\eqref{genus_two_conf_block} is
\begin{equation}
 \mathcal{G}_{\frac{1}{2}, \{\frac{1}{16}, \frac{1}{16}, 0\}}^{(3)}(z)=
 z^{1/72}\left[1+\frac{7}{108}z^+\frac{1595}{46656}z^2+
 \frac{118405}{5038848}z^3+\frac{26160455}{1451188224}z^4+O(z^5)\right].
\end{equation}
Again, when the null vector at level two 
in the Verma module with $c=1/2$ and $h=1/16$ pops up in the sum of Eq.~\eqref{genus_two_conf_block} 
simultaneously in two replicas,  it alters the coefficient of the $O(z^4)$ term. This spurious contribution is again of order $O(\varepsilon^2)$ 
once we apply the regularization of Eq.~\eqref{regularization}, and it is therefore removed in the limit $\varepsilon\to0$.

\section{Orbifold Conformal Blocks in terms of Sphere Conformal Blocks}\label{sec:sphere_conf_blocks}
Eq.~\eqref{conformal_block_decomposition} provides a small $|z|$ expansion of the orbifold conformal block of the type
\begin{equation}
\mathcal{G}_{c,\boldsymbol{h}}^{(N)}(z) =z^{|\boldsymbol{h}|-2h_{\sigma_N}} \;\sum_{j=0}^{\infty} a_j \; z^j.
\end{equation}
However, the computation of the coefficients $a_j$  quickly becomes impossible to accomplish and one has to approximate $\mathcal{G}_{c,\boldsymbol{h}}^{(N)}(z)$ by truncating the previous series at some value $L$,
\begin{equation}\label{trunc}
\mathcal{G}_{c,\boldsymbol{h}}^{(N)}(z) \sim z^{|\boldsymbol{h}|-2h_{\sigma_N}} \;\sum_{j=0}^{L} a_j \; z^j.
\end{equation}
In the case $N=3$, for instance, we are able to reach $L=6$.

A crucial point here is that the convergence of the series in Eq.~\eqref{conformal_block_decomposition} is slow close to  $|z|=1$. This, in turn, implies that, if the conformal block $\mathcal{G}_{c,\boldsymbol{h}}^{(N)}(z)$ is approximated by the truncated sum of Eq.~\eqref{trunc}, one misses the global properties of the twist field four-point function and, in particular, the crossing symmetry of Eq.~\eqref{cross_symmetry}. 
In~\cite{Collier}, this problem was tackled for $N=3$ by using a transformation from the $z$-plane to the pillow frame introduced in~\cite{Maldacena}. Ref.~~\cite{Collier}  derived then a series expansion in terms of the elliptic nome $q(z)=e^{i\pi\tau(z)}$, with $\tau(z)$ defined in Eq.~\eqref{tau}, 
of the form
\begin{equation}\label{collierap}
\mathcal{G}_{c,\boldsymbol{h}}^{(N)}(z)\sim\Lambda(q(z))\sum_{l=0}^{L} A_l \; q(z)^l.
\end{equation}
The explicit expressions of the function $\Lambda(q)$ and of the coefficient $A_l$ are 
given in Sec.~\ref{sec:bound_str_cts}, see Eq.~\eqref{orbifold_cb_pillow}. 
This expansion drastically improves the convergence properties of the twist field 
correlation function near $|z|=1$.

In this section, we will reobtain the expansion in Eq~\eqref{collierap} from a different perspective and show how it could be further improved.
As we discuss in detail in Appendix~\ref{app_orbifold_algebra}, the orbifold algebra admits as a sub-algebra a Virasoro algebra with central charge $Nc$~\cite{Borisov}, generated by the symmetric stress energy tensor in Eq.~\eqref{hatT}. Then it is natural to expand the orbifold conformal blocks 
$\mathcal{G}_{c, \boldsymbol{h}}^{(N)}(z)$ as a linear combination of Virasoro sphere conformal blocks with central charge $Nc$. More specifically,
\begin{equation}\label{expansion_sphere_conf_blocks}
 \mathcal{G}_{c, \boldsymbol{h}}^{(N)}(z)=\sum_{l=0}^\infty\alpha_l^{\boldsymbol{h}}\mathcal{F}_{Nc, |\boldsymbol{h}|+l}(z),
\end{equation}
where we have  denoted by $\mathcal{F}_{Nc, h }(z)$ the Virasoro sphere conformal blocks with the four external dimensions fixed to  $h_{\sigma_N}$ and with internal channel of dimension $h$. The conformal block $\mathcal{F}_{Nc, |\boldsymbol{h}|+l}(z)$ resums the 
contribution of the conformal family generated by a field with conformal dimension $|\boldsymbol{h}|+l$, primary with respect to the 
orbifold Virasoro sub-algebra of Eq.~\eqref{symm_virasoro_alg}. The coefficients $\alpha_l^{\boldsymbol{h}}$ can be thought as Clebsch-Gordan coefficients for a 
decomposition of a $N$-fold tensor product of Virasoro algebra representations into a direct sum of irreducible representations with central charge $Nc$. A more detailed discussion concerning the meaning of Eq.~\eqref{expansion_sphere_conf_blocks}  as well as the algebraic nature of the coefficients $\alpha_{l}^{\boldsymbol{h}}$ can be found in Section~\ref{deralpha}. 

Now observe that, if we truncate Eq.~\eqref{expansion_sphere_conf_blocks}  
\begin{equation}\label{expansion_sphere_conf_blocks_trunc}
 \mathcal{G}_{c, \boldsymbol{h}}^{(N)}(z)\sim\sum_{l=0}^{L}\alpha_{l}^{\boldsymbol{h}}
 \mathcal{F}_{Nc, |\boldsymbol{h}|+l}(z),
\end{equation}
then the coefficients $\alpha_{l}^{\boldsymbol{h}}$ can be easily derived by expanding $\mathcal{F}_{Nc, |\boldsymbol{h}|+l}(z)$ in power series and identifying order by order the sums of Eqs.~\eqref{trunc} and \eqref{expansion_sphere_conf_blocks_trunc}.
Eq.~\eqref{expansion_sphere_conf_blocks_trunc} is already a better approximation than 
Eq.~\eqref{trunc}, as each  $\mathcal{F}_{Nc, |\boldsymbol{h}|+l}(z)$ can be computed until orders much larger than $z^{|\boldsymbol{h}|-2h_{\sigma_N}+L}$. In other words, this means that we can take into account the contribution of certain very low descendants of the orbifold algebra that were previously unaccessible.

The second step is to express the $\mathcal{F}_{Nc, h}(z)$ as an expansion in terms of the elliptic nome $q(z)$ by using the elliptic Zamolodchikov recursion relation~\cite{Zamolodchikov},
\begin{equation}\label{zam_recursion}
 \mathcal{F}_{Nc, h}(z)=f(h, Nc, z) \;H(h, Nc, q(z)),
\end{equation}
where
\begin{align}
f(h, Nc, z) &= [16q(z)]^{h-\frac{Nc-1}{24}}
 [z(1-z)]^{\frac{Nc-1}{24}-2h_{\sigma_N}}
 \vartheta_3(q(z))^{\frac{Nc-1}{2}-16h_{\sigma_N}}, \\
 H(h, Nc, q)&=1+\sum_{j=1}^{\infty}a_j(h, Nc)\;q^{2j}. \label{zam_H}
\end{align}
The coefficients $a_j(h, Nc)$ can be computed recursively to very large values of $j$, see Ref.~\cite{Zamolodchikov}. 

Finally, if the elliptic recursion is truncated at level $L'>L$ and then combined 
with Eq.~\eqref{expansion_sphere_conf_blocks_trunc}, we get the following approximation 
for $\mathcal{G}_{c, \boldsymbol{h}}^{(N)}(z)$, 
\begin{equation}
\label{usap}
\mathcal{G}_{c,\boldsymbol{h}}^{(N)}(z) \sim \Lambda(q) \left(\sum_{l=0}^{L} A_l~q(z)^l+\sum_{l=L+1}^{L+L'} A_l'~q(z)^{l}\right), 
\end{equation}
that provides sub-leading corrections to Eq.~\eqref{collierap}. We distinguish between coefficients 
$A_l$ and $A_l'$ because the former take into account the contribution from all the states at level $l<L$.
 The $q^l$ terms with $l>L$ come from the contribution of the descendants of the primaries at the levels $l<L$ in the orbifold sub-algebra. Therefore, although both Eqs.~\eqref{collierap}~and~\eqref{usap} 
approximate the conformal block $ \mathcal{G}_{c,\boldsymbol{h}}^{(N)}(z)$ with an error $o(q(z)^{L})$, 
Eq.~\eqref{usap} gives a better approximation than Eq.~\eqref{collierap}.  

\subsection{Orbifold Conformal Algebra and the Coefficients $\alpha_l^{\text{\textit{\textbf{h}}}}$}\label{deralpha}

To illustrate the main idea behind Eq.~\eqref{expansion_sphere_conf_blocks}, it is sufficient to consider the first level contribution to  
$\mathcal{G}_{c, \boldsymbol{h}}^{(N)}(z)$ for the case $N=3$. If we explicitly calculate the first order term in the expansion of Eq.~\eqref{genus_two_conf_block}, we have that
\begin{equation}
\mathcal{G}_{c, \boldsymbol{h}}^{(3)}(z)= z^{|\boldsymbol{h}|-2h_{\sigma_N}}\left[1+\left(\frac{|\boldsymbol{h}|}{2}+\frac{1}{54}\left(\frac{(h_1-h_2)^2}{h_3}+\frac{(h_1-h_3)^2}{h_2}+\frac{(h_2-h_3)^2}{h_1}\right)\right)\;z+O(z^2)\right],
\end{equation}
where we recall that here $\boldsymbol{h}=\{h_1,h_2,h_3\}$ and $|\boldsymbol{h}|=h_1+h_2+h_3$. The conformal block $\mathcal{G}_{c, \boldsymbol{h}}^{(3)}(z)$ is associated with the field $\boldsymbol{\phi}_{\boldsymbol{h}}=\phi_{h_1}\otimes\phi_{h_2}\otimes\phi_{h_3}$ and its descendants. In particular, the coefficient of the $z^{|\boldsymbol{h}|-2h_{\sigma_N}+1}$ term comes from the contribution of three descendants,
\begin{equation}
\label{vec_basis}
|v_1\rangle =L_{-1}\phi_{h_1}\otimes \phi_{h_2}\otimes \phi_{h_3},   \quad |v_2\rangle =\phi_{h_1}\otimes  L_{-1}\phi_{h_2}\otimes \phi_{h_3},\quad |v_3\rangle=\phi_{h_1}\otimes \phi_{h_2}\otimes  L_{-1} \phi_{h_3}.
\end{equation}

The symmetric linear combination of the three states above corresponds to the descendant of the orbifold Virasoro sub-algebra
of Eq.~\eqref{symm_virasoro_alg}
\begin{equation}
\boldsymbol{L}_{-1}\boldsymbol{\phi}_{\boldsymbol{h}}=|v_1\rangle+|v_2\rangle+|v_3\rangle.
\end{equation} 
This descendant contributes with the term with coefficient $|\boldsymbol{h}|/2$ to the expansion of Eq.~\eqref{usap}. This contribution is taken into account by the first Virasoro sphere conformal block in the expansion of Eq.~\eqref{expansion_sphere_conf_blocks}, that is 
\begin{equation}
\mathcal{F}_{3c, |\boldsymbol{h}|}(z)=z^{|\boldsymbol{h}|-2h_{\sigma_N}}\left[1+\frac{|\boldsymbol{h}|}{2}z+O(z^2)\right].
\end{equation}

Let us consider now  the orthogonal complement to the span of the vector $\boldsymbol{L}_{-1}\boldsymbol{\phi}_{\boldsymbol{h}}$ in Eq.~\eqref{symm_virasoro_alg}. This is a two dimensional real vector space that contains the states
\begin{equation}
\label{ortosub}
 |\Psi\rangle= \mu|v_1\rangle+\lambda |v_2\rangle
 -\frac{\mu h_1+\lambda h_2}{h_3} |v_3\rangle,
\end{equation}
for $\lambda,\mu\in\mathbb R$.
One can check
that any vector in  the orthogonal complement is a primary of the $3 c$ Virasoro sub-algebra of 
Eq.~\eqref{symm_virasoro_alg}, i.e. $\boldsymbol{L}_n  |\Psi\rangle=0$ for all $n>0$
and $\boldsymbol{L}_0|\Psi\rangle = (h_1+h_2+h_3+1)|\Psi\rangle$.
The contribution of these states and their symmetric descendants $\boldsymbol{L}_{-M}|\Psi\rangle$ 
to the conformal block $\mathcal{G}_{c, \boldsymbol{h}}^{(3)}(z)$ is given by
\begin{equation}
\alpha_{1}^{\boldsymbol{h}}\;\mathcal{F}_{3c, |\boldsymbol{h}|+1}(z),
\end{equation}
where 
\begin{equation}
\label{alpha1}
\alpha_{1}^{\boldsymbol{h}} =\frac{1}{54}\left[\frac{(h_1-h_2)^2}{h_3}+\frac{(h_1-h_3)^2}{h_2}+\frac{(h_2-h_3)^2}{h_1}\right].
\end{equation}
The coefficient $\alpha_{1}^{\boldsymbol{h}}$ can  be determined by choosing an orthogonal basis $\{|\Psi_1\rangle, |\Psi_2\rangle\}$ for the subspace in Eq.~\eqref{ortosub}. Then one can find the result in Eq.\eqref{alpha1} from the expression
\begin{equation}
 \alpha_1^{\boldsymbol{h}}=\frac{1}{D_{\boldsymbol{h},\boldsymbol{\bar{h}}}}
 \left[\frac{\langle \Psi_1\boldsymbol{\phi}_{\boldsymbol{\bar{h}}}|\sigma_{3}(1)|\tilde{\sigma}_3\rangle \;\langle \Psi_1\boldsymbol{\phi}_{\boldsymbol{\bar{h}}}|\tilde{\sigma}_3(1) | \sigma_{3}\rangle}{\langle \Psi_1|\Psi_1\rangle}
 +\frac{\langle \Psi_2\boldsymbol{\phi}_{\boldsymbol{\bar{h}}}|\sigma_{3}(1)|\tilde{\sigma}_3\rangle \;\langle \Psi_2\boldsymbol{\phi}_{\boldsymbol{\bar{h}}}|\tilde{\sigma}_3(1)|\sigma_{3}\rangle}{\langle \Psi_2|\Psi_2\rangle}  \right]
\end{equation}
where $D_{\boldsymbol{h},\boldsymbol{\bar{h}}}$ is related to the structure constant as in Eq.~\eqref{DC}. As we already pointed out for the matrix elements $\rho$ and $\tilde{\rho}$, defined in Eq.~\eqref{Gammas}, the coefficients $\alpha^{\boldsymbol{h}}_1$  are algebraic in nature, despite the holomorphic and anti-holomorphic fields appear in the equation above. This can be seen from the fact that they are rational functions of the central charge $c$ and of the holomorphic dimensions $h_j$.

The states $\boldsymbol{L}_{-M}\boldsymbol{\phi}_{\boldsymbol{h}}$ and $\boldsymbol{L}_{-M} |\Psi_{1,2}\rangle$ 
do not span the full space of descendants of $\boldsymbol{\phi}_{\boldsymbol{h}}$ for levels larger than one. 
This is the reason why other Virasoro conformal blocks appear and one finally gets the expansion of Eq.~\eqref{expansion_sphere_conf_blocks}. 
In fact, the previous analysis may be extended to  higher levels and generalized to any number of copies $N$.
If we denote by $\mathfrak{p}(l)$ the number of partitions of the integer $l$, then there are $\mathfrak{p}(l)$ symmetric
descendants $\boldsymbol{L}_{-M}\boldsymbol{\phi}_{\boldsymbol{h}}$ at level $l$. However,
for an arbitrary number of replicas $N$ and assuming that there are no null vectors, it is easy to check that 
at level $l$ the total number $\mathcal{N}_{N, l}$ of linearly independent descendants of $\boldsymbol{\phi}_{\boldsymbol{h}}$ is
\begin{equation}
 \mathcal{N}_{N, l}=\sum_{\substack{Y, |Y|=l \\ |j_Y|\leq N}}\frac{N!}{(N-|j_Y|)!
 \prod_{i=1}^{|i_Y|}d_Y(i)!}\prod_{j=1}^{|j_Y|}\mathfrak{p}(i_Y(j)),
\end{equation}
where $Y$ denotes a partition of $l$. If we consider the Young tableau
associated to $Y$, then $|i_Y|$ and  $|j_Y|$ denote its number of columns and
rows respectively, $i_Y(j)$ is the number of columns in the row $j$ and 
$d_Y(i)$ is the number of rows with $i$ columns.

Therefore, at level $l$, we shall find a set of $\mathcal{M}_{N,l}$ fields 
$|\Psi_j\rangle$, $j=1,\dots, \mathcal{M}_{N,l}$, orthogonal to all the $\mathfrak{p}(l)$ symmetric descendants at level $l$  and such that $\boldsymbol{L}_0 |\Psi_j\rangle=(|\boldsymbol{h}|+l)|\Psi_j\rangle$
$\boldsymbol{L}_n|\Psi_j\rangle=0$ for all $n>0$. If there are no null vectors, the numbers $\mathcal{M}_{N,l}$ are defined by the recursion relation
\begin{equation}
 \mathcal{M}_{N, l}=\mathcal{N}_{N, l}
 -\sum_{m=0}^{l-1}\mathcal{M}_{N, m}\mathfrak{p}(l-m).
\end{equation}
The contribution of these new  primaries fields, which all have conformal dimension $|\boldsymbol{h}|+l$, and their descendants $\boldsymbol{L}_{-M}|\Psi_j\rangle$  is included into the
sphere conformal block $\mathcal{F}_{Nc, |\boldsymbol{h}|+l}(z)$ in the expansion of the orbifold conformal block, see Eq.~\eqref{expansion_sphere_conf_blocks}. The corresponding structure constant $\alpha_l^{\boldsymbol{h}}$
is given by
\begin{equation}\label{alpha_l}
 \alpha_l^{\boldsymbol{h}}=\frac{1}{D_{\boldsymbol{h},\boldsymbol{\bar{h}}}}\sum_{j=1}^{\mathcal{M}_{N,l}}
 \frac{\langle \Psi_j\boldsymbol{\phi}_{\boldsymbol{\bar{h}}}|\sigma_{N}(1)|\tilde{\sigma}_N\rangle \;\langle \Psi_j\boldsymbol{\phi}_{\boldsymbol{\bar{h}}}|\tilde{\sigma}_N(1) |\sigma_{N}\rangle}
 {\langle \Psi_j|\Psi_j\rangle},
\end{equation}
and can be checked explicitly.
The primary fields at 
different levels, as well as their descendants, are orthogonal
since they belong to different representations of the Virasoro sub-algebra of Eq.~\eqref{symm_virasoro_alg}.
\section{Applications}\label{sec:applications}
We now pass  to discuss applications of the formalism developed in Sec.~\ref{sec:conf_blocks} and Sec.~\ref{sec:sphere_conf_blocks}. In Sec.~\ref{sec:ent},~we will focus on the determination of the R\'enyi entropies and the third power of the partial transpose of the reduced density matrix for two disjoint intervals in a minimal CFT. We further discuss how the expansion in Eq.~\eqref{usap} can be used to improve the bounds on the CFT structure constants  analyzed in~\cite{Collier}, see Sec.~\ref{sec:bound_str_cts} and Sec.~\ref{sec:num_bootstrap}. 
\subsection{Entanglement of Two Disjoint Intervals: R\'enyi Entropies and the Partially Transposed Density Matrix}
\label{sec:ent}
Consider a critical one-dimensional quantum model in the ground state. Let us suppose that 
the system is divided into two spatial regions $A$ and $B$. The entanglement
between these two subsystems can be characterized by the moments $\Tr\rho_A^N$
of the reduced density matrix $\rho_A$, defined in Eq.~\eqref{red_density_matrix} 
of Appendix~\ref{app_ent}. We will further assume that the subsystem $A$ consists 
of two disjoint regions $A_1$ and $A_2$, i.e. $A=A_1\cup A_2$. Then one can also 
study the entanglement between $A_1$ and $A_2$, which can be quantified by the moments 
$\Tr (\rho_A^{T_2})^N$ of the partial transpose of $\rho_A$, which we will denote by $\rho_A^{T_2}$
and defined in Eq.~\eqref{partial_transpose}.

Now we suppose that the universal properties of the quantum model at the critical 
point are determined by the CFT $\mathcal{C}$. Without loss of generality, we can 
take as regions $A_1$ and $A_2$ the intervals $A_1=(0, x)$ and $A_2=(1, \infty)$, 
with $0<x<1$. Then the moments $\Tr \rho_A^N $ and $\Tr (\rho_A^{T_2})^N$ are equal to 
~\cite{Calabrese09, Calabrese11}
 \begin{equation}\label{red_mat_ft}
 \Tr\rho_A^N=K_N\langle \sigma_N(\infty)\tilde{\sigma}_N(1)\sigma_N(x)\tilde{\sigma}_N(0)\rangle,
 \end{equation}
 and~\cite{CalabreseNeg12, CalabreseNeg}
 \begin{equation}\label{partial_trans_ft}
 \Tr(\rho_A^{T_2})^N=K_N\langle \tilde{\sigma}_N(\infty)\sigma_N(1)\sigma_N(x)\tilde{\sigma}_N(0)\rangle,
 \end{equation}
where $K_N$ is a non-universal constant. Note that the partial transposition crucially exchanges the twist and anti-twist operators 
at the points $z=1$ and $\infty$. The crossing transformation $z\mapsto 1-z$ maps subsystem 
$A=(0, x)\cup(1, \infty)$ into its complement $B=(-\infty, 0)\cup(1-x, 1)$. Thus
the crossing invariance of the four-point correlation function $\langle \sigma_N(\infty)
\tilde{\sigma}_N(1) \sigma_N(x) \tilde{\sigma}_N(0)\rangle$ implies that $\Tr\rho_A^N=\Tr\rho_B^N$
and, therefore, the well-known property that for a pure state the entanglement entropies of a 
subsystem and its complementary are equal. This conclusion is no longer true in general for the moments $\Tr(\rho_A^{T_2})^N$
and, in fact, the four-point function $ \langle \tilde{\sigma}_N(\infty)\sigma_N(1)\sigma_N(x)\tilde{\sigma}_N(0)\rangle$
is not crossing invariant. The case $N=2$ is peculiar, since the twist and anti-twist fields  are identified, $\sigma_2=\tilde{\sigma}_2$, and therefore $\Tr\rho_A^2=\Tr(\rho_A^{T_2})^2$.

The correlation functions of Eqs.~\eqref{red_mat_ft} and \eqref{partial_trans_ft} are
related by the conformal transformation
\begin{equation}
 y(z)=\frac{z}{z-1},
\end{equation}
which maps the points $(0, x, 1, \infty)$ into $(0, x/(x-1), \infty, 1)$. We can rewrite then 
\begin{equation}
 \langle \tilde{\sigma}_N(\infty)\sigma_N(1)\sigma_N(x)\tilde{\sigma}_N(0)\rangle=
 (1-x)^{-4h_{\sigma_N}}\langle \sigma_N(\infty)\tilde{\sigma}_N(1)\sigma_N(x/(x-1))\tilde{\sigma}_N(0)\rangle,
\end{equation}
and therefore Eq.~\eqref{partial_trans_ft} can be recast in the form 
\begin{equation}\label{partial_trans_ft_2}
 \Tr(\rho_A^{T_2})^N=K_N(1-x)^{-4h_{\sigma_N}}\langle \sigma_N(\infty)\tilde{\sigma}_N(1)\sigma_N(x/(x-1))\tilde{\sigma}_N(0)\rangle.
 \end{equation}
Thus, while for determining $\Tr\rho_A^N$ it is enough to calculate the twist field four-point correlator
in the interval $0<x<1$,  the computation of $\Tr(\rho_A^{T_2})^N$ requires to extend this function to the domain $x/(x-1)<0$. The moments of the reduced density matrix can be calculated directly from  Eqs.~\eqref{conformal_block_decomposition_0}
and \eqref{expansion_sphere_conf_blocks}, and we have
\begin{equation}\label{red_mat_cb}
 \Tr\rho_A^N=K_N\sum_{\boldsymbol{h}, \boldsymbol{\bar{h}}}\sum_{l, l'} 
 \tilde{D}_{\boldsymbol{h}, \boldsymbol{\bar{h}}}^{l, l'} 
 \mathcal{F}_{Nc, |\boldsymbol{h}|+l}(x)\mathcal{F}_{Nc, |\boldsymbol{\bar{h}}|+l'}(x)
\end{equation}
with $\tilde{D}_{\boldsymbol{h}, \boldsymbol{\bar{h}}}^{l, l'}=D_{\boldsymbol{h}, 
\boldsymbol{\bar{h}}}\alpha_l^{\boldsymbol{h}}\alpha_{l'}^{\boldsymbol{\bar{h}}}.$
As previously emphasized, in order to obtain an analogous expansion for the moments of the partially transposed density matrix one must determine the analytic continuation of the twist field four-point function along 
the negative real axis. Such analytic continuation can be straightforwardly calculated from the expression of the orbifold conformal 
blocks in terms of the elliptic nome $q(x)=e^{i\pi \tau(x)}$, see Eq.~\eqref{usap}. Indeed, one can easily prove that
\begin{equation}\label{tau_p_1}
 \tau\left(\frac{x}{x-1}\right)=\tau(x)+1, \quad \text{and} \quad 
 q\left(\frac{x}{x-1}\right)=e^{i\pi} q(x).
\end{equation}
In other words, $\Tr(\rho_A^{T_2})^N$ can  be again computed from Eqs.~\eqref{conformal_block_decomposition_0} 
and \eqref{expansion_sphere_conf_blocks} but evaluating now
the Virasoro conformal blocks at $e^{i\pi}q(x)$,
\begin{equation}\label{partial_trans_cb}
 \Tr(\rho_A^{T_2})^N=K_N(1-x)^{-4h_{\sigma_N}}\sum_{\boldsymbol{h}, \boldsymbol{\bar{h}}}\sum_{l, l'} 
 \tilde{D}_{\boldsymbol{h}, \boldsymbol{\bar{h}}}^{l, l'} 
 \mathcal{F}_{Nc, |\boldsymbol{h}|+l}(e^{i\pi}q)\mathcal{F}_{Nc, |\boldsymbol{\bar{h}}|+l'}(e^{-i\pi}q).
\end{equation}
Note that, for convenience,  in Eq.~\eqref{partial_trans_cb} and some more equations below, we have traded $x$ for the elliptic nome $q$ 
in the argument of $\mathcal{F}_{Nc, h}$. Observe that, for $N=2$, Eq.~\eqref{tau_p_1} corresponds to perform the modular 
transformation $\tau(x)\mapsto \tau(x)+1$ on the modulus of the torus $\Sigma_1(x)$.
Then $\mathcal{Z}_1(x)$ and $\mathcal{Z}_1(x/(x-1))$ are the partition functions
of $\mathcal{C}$ on a flat torus of moduli $\tau(x)$ and $\tau(x)+1$ respectively. Taking into account Eqs.
\eqref{red_mat_ft} and \eqref{partial_trans_ft_2}, the invariance of the partition function on the torus under modular transformations, 
$\mathcal{Z}_1(x)=\mathcal{Z}_1(x/(x-1))$, implies the identity $\Tr\rho_A^2=\Tr(\rho_A^{T_2})^2$ anticipated earlier.

We can recast Eq.~\eqref{partial_trans_cb} in a simpler and instructive form. By recalling the 
Zamolodchikov recursion for the sphere conformal blocks and  applying 
the  identities of the elliptic functions
\begin{equation}
 \vartheta_2(e^{\pm i\pi}q)=e^{\pm i\pi/4}\vartheta_2(q), \quad 
 \vartheta_3(e^{\pm i\pi}q)=\vartheta_4(q), \quad 
 \vartheta_4(e^{\pm i\pi}q)=\vartheta_3(q),
\end{equation}
and 
\begin{equation}
 x=\left(\frac{\vartheta_2(q)}{\vartheta_3(q)}\right)^4, \quad 
 1-x=\left(\frac{\vartheta_4(q)}{\vartheta_3(q)}\right)^4,
\end{equation}
one can show that 
\begin{equation}
 \mathcal{F}_{Nc, h}(e^{\pm i\pi}q)=e^{\pm i\pi(h-2h_{\sigma_N})}
 (1-x)^{2h_{\sigma_N}}\mathcal{F}_{Nc, h}(q).
\end{equation}
By plugging the last equality into Eq.~\eqref{partial_trans_cb}, the conformal block
expansion of the moments $\Tr(\rho_A^{T_2})^N$ can be eventually rewritten
as
\begin{equation}\label{partial_trans_cb_2}
 \Tr(\rho_A^{T_2})^N=K_N\sum_{\boldsymbol{h}, \boldsymbol{\bar{h}}}\sum_{l, l'} 
 e^{i\pi(|\boldsymbol{h}|-|\boldsymbol{\bar{h}}|+l-l')}\tilde{D}_{\boldsymbol{h}, \boldsymbol{\bar{h}}}^{l, l'} 
 \mathcal{F}_{Nc, |\boldsymbol{h}|+l}(x)\mathcal{F}_{Nc, |\boldsymbol{\bar{h}}|+l'}(x).
\end{equation}  
By comparing Eq.~\eqref{partial_trans_cb_2} above with Eq.~\eqref{red_mat_cb}, we conclude that the moments $\Tr(\rho_A^{T_2})^N$
admit the same Virasoro conformal block decomposition as $\Tr\rho_A^N$, 
but with the structure constants multiplied by a spin dependent phase.
We can then state the main result of this section as
\begin{equation}
 \Tr\rho_A^N-\Tr(\rho_A^{T_2})^N=\text{sum over the channels with conformal spin}\,\,
 |\boldsymbol{h}|-|\boldsymbol{\bar{h}}|+l-l'\neq 2k,\quad k\in\mathbb{Z}.
\end{equation}
Note that, in the case $N=2$, due 
to the identity $\Tr\rho_A^2=\Tr(\rho_A^{T_2})^2$, the channels indicated above cannot appear in the 
conformal block decompositions of $\Tr\rho_A^2$ and $\Tr(\rho_A^{T_2})^2$.

To the best of our knowledge, for $c\leq 1$, analytic expressions of $\Tr\rho_A^N$ and $\Tr(\rho_A^{T_2})^N$ have  
only been calculated  for free theories; namely  the compactified massless boson, the massless Dirac and Majorana fermions \cite{Furukawa, Calabrese09, Calabrese11, CalabreseNeg12, CalabreseNeg, Tagliacozzo, Alba2, Alba3}.  The latter corresponds to the Ising CFT, which is a minimal model with $c=1/2$, and we shall focus on it first. For
comparing with the previous literature, let us rewrite the moments $\Tr\rho_A^N$ 
and $\Tr(\rho_A^{T_2})^N$ in the form~\cite{Calabrese09}
\begin{equation}\label{F_n_cc}
\Tr\rho_A^N=K_Nx^{-4h_{\sigma_N}}(1-x)^{-4h_{\sigma_N}}\mathcal{R}_N(x)
\end{equation}
and~\cite{CalabreseNeg}
\begin{equation}\label{G_n_cc}
 \Tr(\rho_A^{T_2})^N=K_N x^{-4h_{\sigma_N}}(1-x)^{4h_{\sigma_N}}\mathcal{R}_N\left(\frac{x}{x-1}\right).
\end{equation}
In Refs.~\cite{Calabrese11, Tagliacozzo}, an exact expression for the function
$\mathcal{R}_N(z)$, $z\in\mathbb{C}$, was found in the Ising CFT, 
which is also reported in Eq.~\eqref{R_n_ising} of Appendix~\ref{app_ising}. 
We have checked that, for $N=2,3$, Eq.~\eqref{R_n_ising} is exactly reproduced in the intervals $0<x<1$ and $x/(x-1)<0$ by the 
conformal block expansions of Eqs.~\eqref{red_mat_cb} and \eqref{partial_trans_cb_2} 
respectively. In Appendix~\ref{app_ising}, we report the explicit 
conformal block decomposition, with the values of the structure constants, 
of the twist field four-point function in the Ising CFT for $N=3$ up to fifth order. 

\begin{figure}[t]
 \begin{minipage}{0.5\linewidth}
 \centering 
 \includegraphics[width=\textwidth]{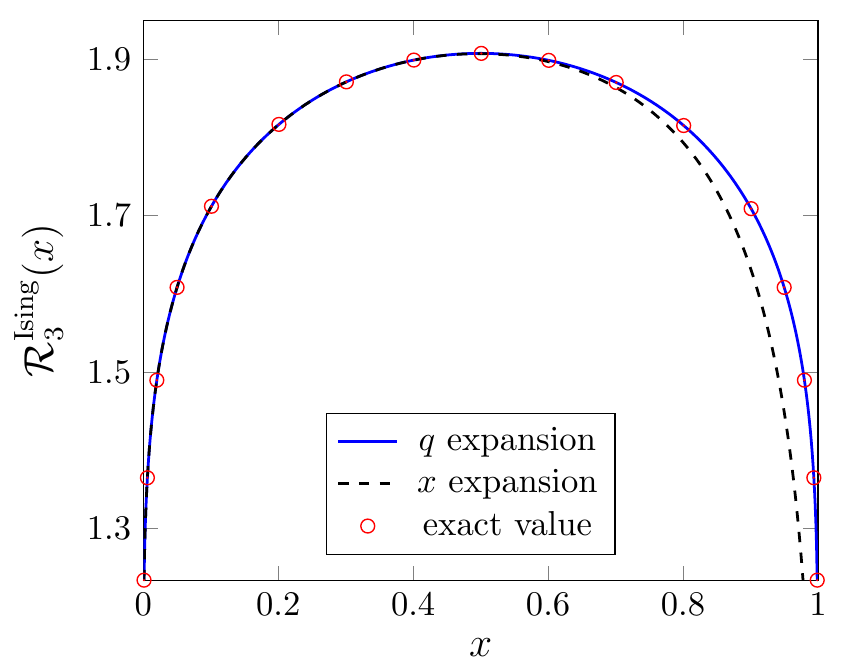}
\end{minipage}
 \begin{minipage}{0.5\linewidth}
 \centering 
 \includegraphics[width=\textwidth]{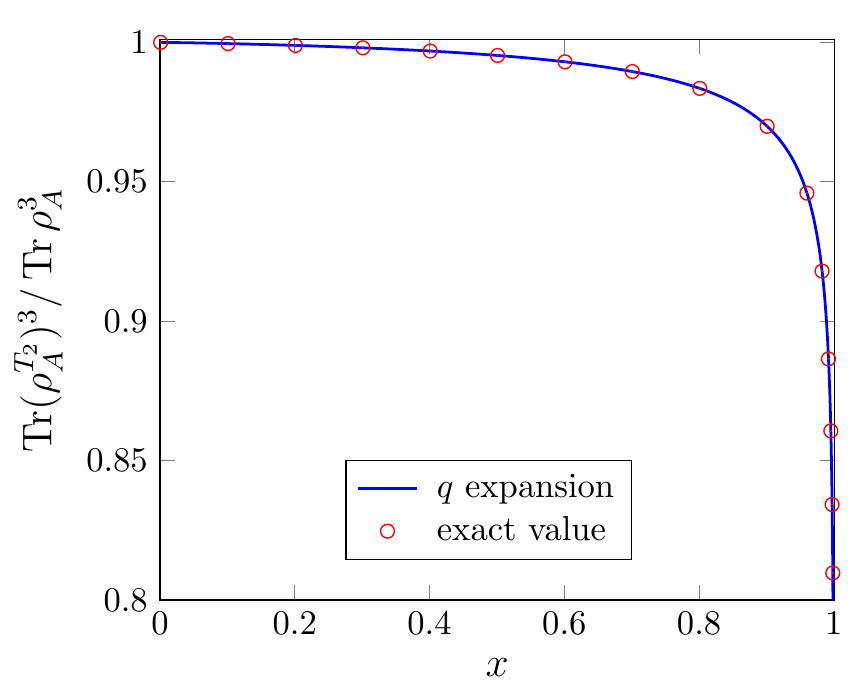}
\end{minipage}
\caption{Analysis of the moments $\Tr\rho_A^N$ and $\Tr(\rho_A^{T_2})^N$ for $N=3$ in the 
free Majorana fermion (Ising CFT). In the left panel, we consider the function $\mathcal{R}_3(x)$ defined in
Eq.~\eqref{F_n_cc}. The dots correspond to the exact values 
calculated using Eq.~\eqref{R_n_ising}. The continuous 
line has been obtained by approximating, in Eq.~\eqref{F_n_cc}, $\Tr\rho_A^3$ with its expansion in  sphere conformal blocks, see Eq.~\eqref{red_mat_cb} and Appendix~\ref{app_ising}, truncated
at level $L=6$. The dashed curve has been determined instead by decomposing $\Tr\rho_A^3$ in terms 
of $N=3$ orbifold conformal blocks, see Eqs.~\eqref{red_mat_ft} and \eqref{orbifold_cb_ising}, and approximating these functions with their  power 
series about $x=0$ in Eq.~\eqref{genus_two_conf_block} up to $L=6$. In the right panel, we study the ratio
$\Tr(\rho_A^{T_2})^3/\Tr\rho_A^3$. The dots are the exact values computed from Eqs.~\eqref{F_n_cc} and 
\eqref{G_n_cc} by substituting Eq.~\eqref{R_n_ising} for the function $\mathcal{R}_3$. The continuous 
line is the result if we approximate  $\Tr\rho_A^3$ and $\Tr(\rho_A^{T_2})^3$ with their expansions in terms of
sphere conformal blocks up to level $L=6$. In all the cases, the sphere conformal blocks have been
calculated using the elliptic recursion of Eq.~\eqref{zam_recursion}, truncated at level $L'=8$.}\label{fig:ent_ising}
\end{figure}

In Fig.~\ref{fig:ent_ising} left, we plot the function $\mathcal{R}_3^{\text{Ising}}(x)$, see Eq.~\eqref{F_n_cc}. 
The continuous line has been drawn by employing the regularized conformal block expansion of Eq.~\eqref{red_mat_cb} 
for $\Tr\rho_A^3$ and the dots are the exact values given by Eq.~\eqref{R_n_ising}. 
The agreement between both results is excellent. Note that, due to the crossing symmetry of the 
twist field four-point function and, therefore, of $\Tr\rho_A^N$, $\mathcal{R}_N(x)$ must 
satisfy in general that $\mathcal{R}_N(x)=\mathcal{R}_N(1-x)$. On the other hand, if we 
decompose $\Tr\rho_A^3$ in $N=3$ orbifold conformal blocks and we expand them using the 
small $x$ representation given in Eq.~\eqref{genus_two_conf_block}, then the result is the dashed curve, which clearly does not display this symmetry for $x$ close to one. In Fig.~\ref{fig:ent_ising} right, we plot the quotient $\Tr(\rho_A^{T_2})^3/\Tr\rho_A^3$. 
The solid curve has been obtained by calculating the conformal block expansions of Eq.~\eqref{red_mat_cb} 
and \eqref{partial_trans_cb_2} for $\Tr\rho_A^3$ and $\Tr(\rho_A^{T_2})^3$ while 
the dots represents the ratio of Eqs.~\eqref{F_n_cc} and \eqref{G_n_cc}, computed by applying Eq.~\eqref{R_n_ising}. Again the conformal block
expansion matches with the previously known results.

In Refs.~\cite{Rajabpour, Ruggiero}, the function $\mathcal{R}_3(x)$ for the Ising CFT was
already studied by expanding the twist field four-point function in sphere conformal
blocks and employing the Zamolodchikov elliptic recursion to speed up its convergence rate. However, in our formalism, the mentioned works only took into account in Eq.~\eqref{red_mat_cb} contributions from the Virasoro sub-algebra primaries at the level $l=0$.
As explained in Sec.~\ref{sec:sphere_conf_blocks}, the inclusion of the orbifold descendants in  Eq.~\eqref{genus_two_conf_block} permits to keep track instead of all the Virasoro sub-algebra primaries up to level $l=6$ in Eq.~\eqref{expansion_sphere_conf_blocks_trunc}. By reaching
higher levels in the conformal block expansion,  one obviously gets a much better cross-symmetric approximation for the twist field correlator, 
cf. Fig. 2 of Ref.~\cite{Ruggiero}. Moreover, a complete understanding of the conformal block expansion of the twist field four-point function allowed us to characterize field theoretically, see Eq.~\eqref{partial_trans_cb_2}, the difference between the two partial traces $\Tr(\rho_A^{T_2})^3$ and $\Tr\rho_A^3$.

\begin{figure}[t]
 \begin{minipage}{0.5\linewidth}
 \centering 
 \includegraphics[width=\textwidth]{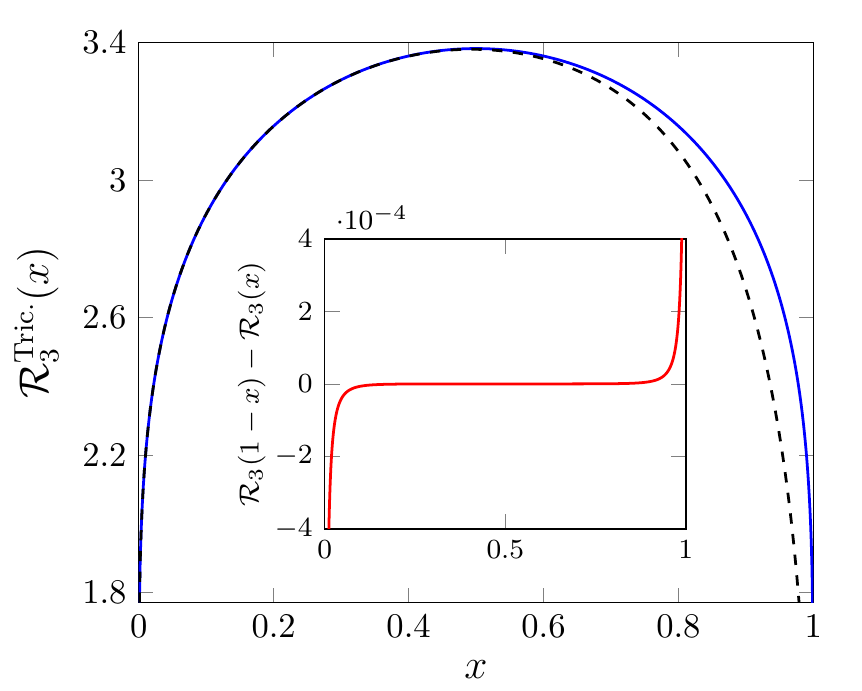}
\end{minipage}
 \begin{minipage}{0.5\linewidth}
 \centering 
 \includegraphics[width=\textwidth]{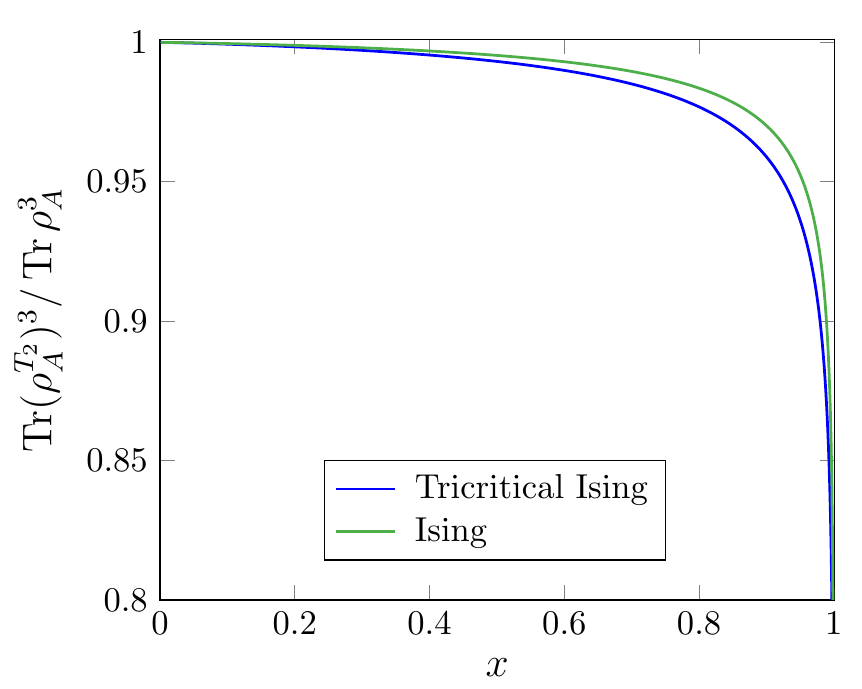}
\end{minipage}
\caption{Analysis of the moments $\Tr\rho_A^N$ and $\Tr(\rho_A^{T_2})^N$ for $N=3$ in the Tricritical 
Ising CFT. In the left panel, we plot the function $\mathcal{R}_3(x)$, introduced in Eq.~\eqref{F_n_cc}, by 
replacing $\Tr\rho_A^3$ with its expansion in sphere conformal blocks, Eq.~\eqref{red_mat_cb}, 
which we truncate at level $L=6$. On the other hand, the dashed line has been computed by decomposing $\Tr\rho_A^3$ in 
$N=3$ orbifold conformal blocks and substituting their small $x$ representation in Eq.~\eqref{genus_two_conf_block}. In the 
inset of left panel, we check the crossing symmetry of the continuous blue line. In the right panel, 
we study the ratio $\Tr(\rho_A^{T_2})^3/\Tr\rho_A^3$. The curve has been obtained by applying 
the sphere conformal block expansions for $\Tr\rho_A^3$ and $\Tr(\rho_A^{T_2})^3$ of Eqs. \eqref{red_mat_cb} and \eqref{partial_trans_cb_2}, 
up to level $L=6$. As a comparison, we also plot the result presented in Fig.~\ref{fig:ent_ising} for the Ising CFT.
The sphere conformal blocks have been computed until level $L'= 8$ using the elliptic recursion of Eq.~\eqref{zam_recursion}.}\label{fig:ent_tricritical}

\end{figure}

Finally, as a further application, in Fig.~\ref{fig:ent_tricritical}, we plot the results for the moments $\Tr\rho_A^3$ and 
$\Tr(\rho_A^{T_2})^3$ in the Tricritical Ising CFT with $c=7/10$ (see~\cite{Mussardo}). The Tricritical Ising CFT appears at the quantum critical point of a device for quantum computation known as the golden chain~\cite{Feiguin}. It is an interacting CFT for which, especially concerning entanglement of disjoint intervals, there are no 
previous results available in the literature. In Fig.~\ref{fig:ent_tricritical} left, we plot 
the function $\mathcal{R}_3(x)$ for this theory, extracted from the corresponding conformal block 
decomposition of $\Tr\rho_A^3$. In Appendix~\ref{app_tric_ising}, one can find the explicit conformal block expansion, 
as well as the values of the structure constants, of the twist field four-point function for the 
Tricritical Ising CFT when $N=3$ up to fifth order. In the inset, we check the crossing invariance
of $\mathcal{R}_3^{\text{Tric.}}(x)$ using the sphere conformal block expansion. We observe that the function is 
symmetric under $x\mapsto 1-x$ for almost all $x\in(0,1)$. In the same plot, we  also included the curve (the dashed line)
that one determines by decomposing $\Tr\rho_A^3$  in $N=3$ orbifold conformal blocks and employing the
small $x$ representation of Eq.~\eqref{genus_two_conf_block}.  As it is clear from the plot, the result 
fails to be cross-symmetric.
In Fig.~\ref{fig:ent_tricritical} right, we plot the ratio
of moments $\Tr(\rho_A^{T_2})^3/\Tr\rho_A^3$ determined  from the conformal block decomposition of Eqs.~\eqref{red_mat_cb} and \eqref{partial_trans_cb_2}.  Here we also compare with the results found for the Ising CFT, the green line.

\subsection{Constraints on the CFT Structure Constants}
\label{sec:bound_str_cts}

In Ref.~\cite{Collier}, the crossing symmetry of the twist field four-point 
function for the case $N=3$ was employed to extract some non-trivial constraints on the 
structure constants of the seed theory $\mathcal{C}$. For this purpose, the authors mapped 
the orbifold conformal block $\mathcal{G}_{c, \boldsymbol{h}}^{(3)}(z)$ from the 
$z$-plane to the 3-fold pillow frame, which makes apparent some positivity properties of 
the block for unitary theories. As we already mentioned in Sec.~\ref{sec:sphere_conf_blocks}, this transformation
produces a series for the orbifold conformal block in terms of the elliptic nome $q(z)$
of the form of Eq.~\eqref{collierap}. In this section, we will see how to rederive 
such constraints on the structure constants from the expansion of the orbifold 
conformal blocks in terms of the ones on the sphere, studied in Sec.~\ref{sec:sphere_conf_blocks}. This 
expansion results in a power series in $q(z)$ of the form of Eq.~\eqref{usap}
Therefore, as we will see, it provides slightly tighter constraints on the structure
constants than the expansion used in Ref.~\cite{Collier}.

First, as in Ref.~\cite{Collier}, we must recast the Zamolodchikov recursion relation, 
Eq.~\eqref{zam_recursion}, for the Virasoro conformal blocks $\mathcal{F}_{Nc, h}(z)$ 
in order to make manifest certain positivity properties  that we will need later.  
In Ref.~\cite{Maldacena}, the conformal block $\mathcal{F}_{Nc, h}(z)$, which is defined 
on the sphere, is mapped to the pillow (the quotient of a flat torus by $\mathbb{Z}_2$, 
which is topologically equivalent to a sphere with four conical singularities). The 
important result for us is that the transformed blocks $\tilde{\mathcal{F}}_{Nc, h}(z)$ read 
\begin{equation}\label{Vir_cb_trans_pill}
 \tilde{\mathcal{F}}_{Nc, h}(q)=\mathcal{\vartheta}_3(q)^{16h_{\sigma_N}-\frac{Nc}{2}}
 \left[z(1-z)\right]^{2h_{\sigma_N}-\frac{Nc}{24}}\mathcal{F}_{Nc, h}(z).
\end{equation}
If $\tilde{\mathcal{F}}_{Nc, h}(z)$ is interpreted as a sum over the states on the pillow, 
then it admits the expansion
\begin{equation}
 \tilde{\mathcal{F}}_{Nc, h}(q)=q^{h-\frac{Nc}{24}}\sum_{n=0}^\infty \tilde{a}_n(h, Nc) q^{n},
\end{equation}
where the coefficients $\tilde{a}_n(h, Nc)$ are sums of scalar products between the descendant 
states at level $n$. Consequently, they are non-negative, $\tilde{a}_n(h, Nc)\geq 0$, for 
unitary theories. On the other hand, if the elliptic recursion of Eq.~\eqref{zam_recursion} is applied in 
Eq.~\eqref{Vir_cb_trans_pill}, one concludes that 
\begin{equation}
 \tilde{\mathcal{F}}_{Nc, h}(q)=
 (16q)^{h-\frac{Nc}{24}}
 \prod_{k=1}^\infty\left(1-q^{2k}\right)^{-\frac{1}{2}}H(h, Nc, q).
\end{equation}
This identity implies that the coefficients $\tilde{a}_n(h, Nc)$ are strictly 
related to the coefficients $a_j(h, Nc)$, which in general are not positive definite, 
of the expansion of $H(h, Nc, q)$, see Eq.~\eqref{zam_H}. In conclusion, we can rewrite the 
Virasoro conformal blocks in terms of the (positive definite) coefficients $\tilde{a}_n(h, Nc)$,
\begin{equation}\label{Vir_cb_pillow}
 \mathcal{F}_{Nc, h}(z)=q^{h-\frac{Nc}{24}}\left[z(1-z)\right]^{\frac{Nc}{24}-2h_{\sigma_N}}
 \vartheta_3(q)^{\frac{Nc}{2}-16h_{\sigma_N}}\sum_{n=0}^\infty \tilde{a}_n(h, Nc) q^n.
\end{equation}

If we now truncate at order $L'>L$ the power series of Eq.~\eqref{Vir_cb_pillow} and insert it into the truncated
expansion of $\mathcal{G}_{c, \boldsymbol{h}}(z)$ in terms of $\mathcal{F}_{Nc, |\boldsymbol{h}|+l}(z)$
of Eq.~\eqref{expansion_sphere_conf_blocks_trunc}, we get the following approximation for the orbifold 
conformal blocks, cf. Eqs.~\eqref{collierap} and \eqref{usap},
\begin{equation}\label{orbifold_cb_pillow}
 \mathcal{G}_{c, \boldsymbol{h}}^{(N)}(z)
 \sim q^{|\boldsymbol{h}|-\frac{Nc}{24}}\left[z(1-z)\right]^{\frac{Nc}{24}-2h_{\sigma_N}}
 \vartheta_3(q)^{\frac{Nc}{2}-16h_{\sigma_N}}
 \left(\sum_{l=0}^L A_l q^l+\sum_{l=L+1}^{L+L'} A_l' q^l\right).
\end{equation}
The coefficients $A_l$ and $A_l'$ are sums of terms of the form 
$\alpha_m^{\boldsymbol{h}} \tilde{a}_{n}(|\boldsymbol{h}|+m, Nc)$,
as one can easily conclude from the combination of Eqs.~\eqref{expansion_sphere_conf_blocks_trunc}~and~\eqref{Vir_cb_pillow}. 
The structure constants $\alpha_m^{\boldsymbol{h}}$ are, by their definition 
in Eq.~\eqref{alpha_l}, non-negative in unitary theories. This implies that 
both $A_l$ and $A_l'$ are also non-negative for any $l$. If we only consider 
the term with coefficients $A_l$, the approximation of Eq.~\eqref{orbifold_cb_pillow} 
reduces to the one used in Ref.~\cite{Collier} for the case $N=3$. As we already remarked 
in Sec.~\ref{sec:sphere_conf_blocks}, the expansion in Virasoro conformal blocks of 
Eq.~\eqref{expansion_sphere_conf_blocks_trunc}, together with the elliptic recursion 
of Eq.~\eqref{zam_recursion}, allows to further incorporate the contribution of some 
of the descendant states at higher levels, the term with coefficients $A_l'$ in 
Eq.~\eqref{orbifold_cb_pillow}, improving the approximation of Ref.~\cite{Collier}.

We can now apply the result of Eq.~\eqref{orbifold_cb_pillow} to derive 
some constraints on the structure constants of the seed theory $\mathcal{C}$.
Using the decomposition of Eq.~\eqref{conformal_block_decomposition_0} of the twist field correlation function 
in terms of orbifold conformal blocks, we can rewrite the crossing symmetry condition 
of Eq.~\eqref{cross_symmetry} in the form
\begin{equation}\label{cross_symmetry_1}
 \sum_{\boldsymbol{h}, \boldsymbol{\bar{h}}} D_{\boldsymbol{h}, \boldsymbol{\bar{h}}}\left[
 \mathcal{G}_{c, \boldsymbol{h}}^{(N)}(z)\mathcal{G}_{c, \boldsymbol{\bar{h}}}^{(N)}(\bar{z})-
 \mathcal{G}_{c, \boldsymbol{h}}^{(N)}(1-z)\mathcal{G}_{c, \boldsymbol{\bar{h}}}^{(N)}(1-\bar{z})\right]=0.
\end{equation}
In the rest of this section, we will restrict to the case $N=3$, 
for which the coefficients $D_{\boldsymbol{h}, \boldsymbol{\bar{h}}}$ are, according 
to Eq.~\eqref{D_3}, proportional to the square of the structure constants 
$C_{\boldsymbol{h}, \boldsymbol{\bar{h}}}^{{\rm seed}}$ of the seed 
theory $\mathcal{C}$. If, as in the usual numerical bootstrap approach 
\cite{Rattazzi, Poland}, we act on Eq.~\eqref{cross_symmetry_1} with a linear functional
\begin{equation}
\label{linfunder}
 \gamma(f)=\sum_{n, m} \gamma_{n, m}\partial_{z}^n\partial_{\bar{z}}^m f(z,\bar{z})\Bigr|_{z=\bar{z}=\frac{1}{2}},
\end{equation}
where $\gamma_{n, m}$ are real coefficients and $f(z,\bar{z})$ an arbitrary function, then 
we can derive a set of linear equations for $(C_{\boldsymbol{h}, \boldsymbol{\bar{h}}}^{{\rm seed}})^2$.
To compare with the results of Ref.~\cite{Collier}, let us take the linear functional that only 
contains the first derivative, $\gamma\equiv\partial_z|_{z=\bar{z}=\frac{1}{2}}$. If we apply it to 
Eq.~\eqref{cross_symmetry_1} for $N=3$, we find the condition
\begin{equation}\label{cross_symm_N_3}
 \sum_{\boldsymbol{h},\boldsymbol{\bar{h}}} 27^{-|\boldsymbol{h}|-|\boldsymbol{\bar{h}}|}
 (C_{\boldsymbol{h}, \boldsymbol{\bar{h}}}^{{\rm seed}})^2~\mathcal{G}_{\boldsymbol{\bar{h}}}^{(3)}\left(1/2\right)
 \partial_z\mathcal{G}_{\boldsymbol{h}}^{(3)}(z)\Bigr|_{z=\frac{1}{2}}=0.
\end{equation}
For unitary theories, $(C_{\boldsymbol{h}, \boldsymbol{\bar{h}}}^{{\rm seed}})^2$ are positive and, 
in order Eq.~\eqref{cross_symm_N_3} to  be satisfied, the factor $\mathcal{G}_{\boldsymbol{\bar{h}}}^{(3)}(z)
\partial_z\mathcal{G}_{\boldsymbol{h}}^{(3)}(z)\bigr|_{z=1/2}$ must be negative on a domain $\mathscr{D}$
of the space of triplets of conformal dimensions $\{(h_1, h_2, h_3)\in \mathbb{R}^3~|~h_1, h_2, h_3\geq 0\}$,
and non-negative otherwise. The points $(h_1, h_2, h_3)$ where this factor vanishes are the boundary
of the domain $\mathscr{D}$ and typically form a compact surface. Thus Eq.~\eqref{cross_symm_N_3} implies that the structure constants 
corresponding to points outside $\mathscr{D}$ are bounded by those associated to points inside it. 

\begin{figure}[t]
 \centering 
 \includegraphics[width=\textwidth]{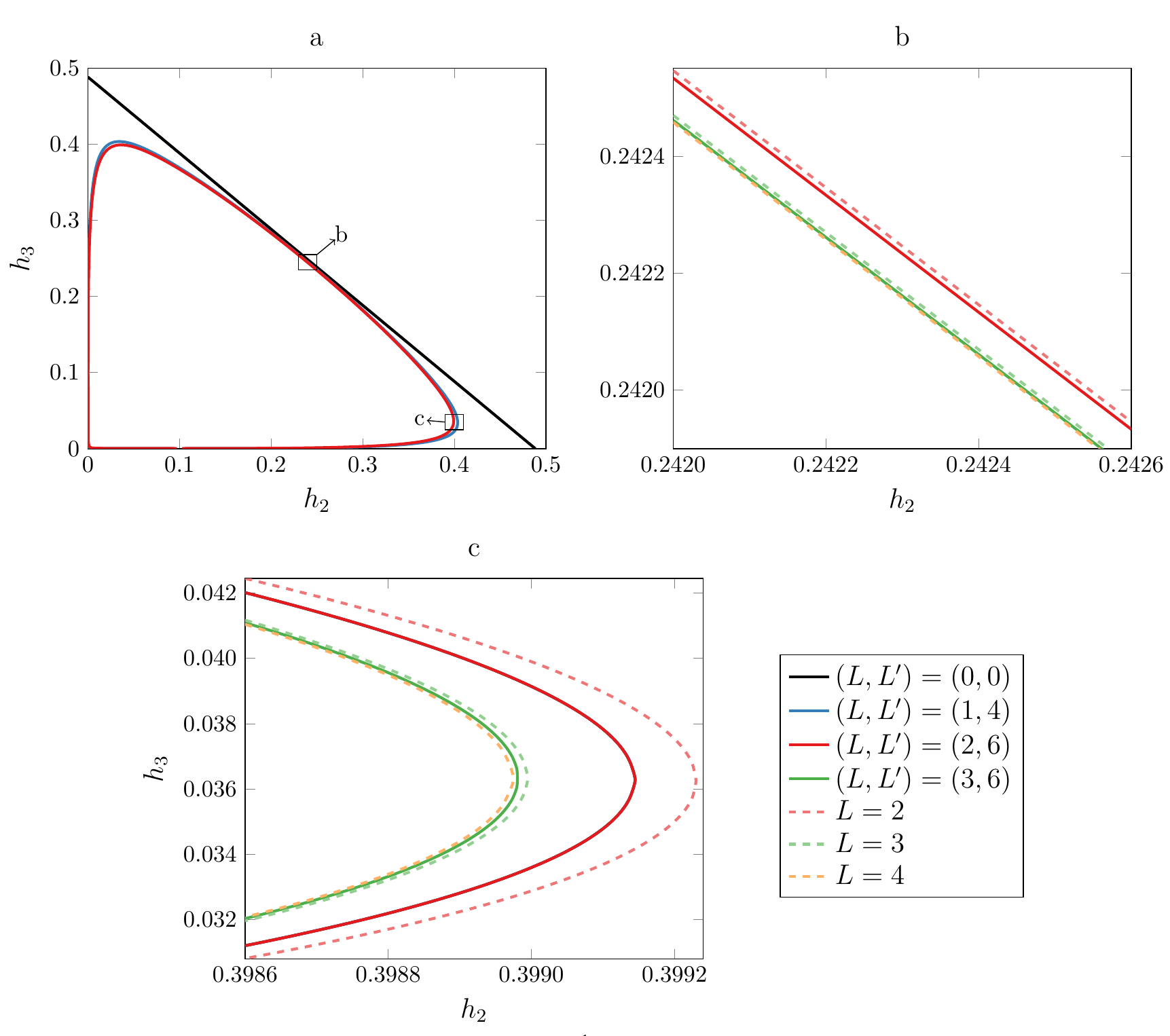}
\caption{The continuous lines represent the points in the $(h_2, h_3)$ plane that 
satisfy the equation $W_c^{(L, L')}(h_1, h_2, h_3)=0$ for $c=4$ and $h_1=0.1$ and several 
values of the truncation levels $L, L'$, see Eq.~\eqref{crit_surf_trunc}. The dashed lines 
are the points that satisfy the same equation but removing from $W_c^{(L, L')}(h_1, h_2, h_3)$
the sums with coefficients $A_l'$. This is the approximation considered in Ref.~\cite{Collier}.
The colouring of the dashed lines corresponds to take different upper bounds $L$
in the sums with coefficients $A_l$. The panels b and c are the magnification of the regions
indicated by a box in the plot a.}\label{fig:constraints_strt}
\end{figure}

The condition of being a point in the boundary of  the domain $\mathscr{D}$
can be rewritten as
\begin{equation}
 W_c(h_1, h_2, h_3)=0,\quad \text{with}\quad 
 W_c(h_1, h_2, h_3)\equiv\partial_z\log\mathcal{G}_{\boldsymbol{h}}^{(3)}(z)\Bigr|_{z=\frac{1}{2}}.
\end{equation}
If we now apply the approximation found in Eq.~\eqref{orbifold_cb_pillow} for 
$\mathcal{G}_{\boldsymbol{h}}^{(3)}(z)$, then we find
\begin{multline}\label{crit_surf_trunc}
 W_c^{(L, L')}(h_1, h_2, h_3)=
 \frac{\pi^2}{K(\frac{1}{2})^2}
 \left[h_1+h_2+h_3-\left(\frac{1}{8}+\frac{5}{72\pi}\right)c\right. \\
 \left.+\frac{\sum_{l=1}^L l A_l e^{-\pi l}+\sum_{l=L+1}^{L+L'} l A_l'e^{-\pi l}}
 {\sum_{l=0}^L A_l e^{-\pi l}+\sum_{l=L+1}^{L+L'} A_l' e^{-\pi l}}\right]. 
\end{multline}
Observe that it is at this point that the positivity of the coefficients $A_l$ 
and $A_l'$ for unitary theories previously discussed plays the crucial role, since 
it implies that the last term in Eq.~\eqref{crit_surf_trunc} is positive too. This means that the 
domain $\mathscr{D}_{L, L'}$ of points $(h_1, h_2, h_3)$ for which $W_c^{(L, L')}(h_1, h_2, h_3)<0$
shrinks as $L$ and $L'$ increase and it converges to the domain $\mathscr{D}$, $\mathscr{D}_{L, L'}\to\mathscr{D}$,
in the limit $L, L'\to\infty$. If in the last term we only include the sum over the coefficients 
$A_l$, we recover the approximation considered in Ref.~\cite{Collier} (cf. Eq. (3.8) of that reference). 
By including the contribution of some of the descendants of the sub-algebra primary fields at 
the levels $l=1, \dots, L$, we improve the convergence of Eq.~\eqref{crit_surf_trunc} and produce a slightly smaller domain 
$\mathscr{D}_{L, L'}$, as Fig.~\ref{fig:constraints_strt} shows. In any case, the convergence with the domain $\mathscr{D}$ is 
very fast due to the exponential decay of those terms in Eq.~\eqref{crit_surf_trunc}.

\subsection{Bootstrapping Genus Two Partition Functions}\label{sec:num_bootstrap}

In this final section, we implement  a numerical bootstrap approach based on Virasoro sphere conformal blocks
to determine numerically the structure constants $D_{\boldsymbol{h}, \boldsymbol{\bar{h}}}$  
of Eq.~\eqref{conformal_block_decomposition_0} for $N=2$ and $N=3$. For minimal seed theories and $N=2,3$,  these constants are known, see Eq.\eqref{D_2} and Eq.~\eqref{D_3}. However, the  motivation of this section is twofold. First, we show that, even in the simplest case, the stability of the numerical procedure  is greatly sensitive to the symmetry $x\mapsto 1-x$ of the orbifold conformal blocks. As a matter of fact,  the numerical outcomes start to converge only if one uses the approximation of Eq.~\eqref{usap}. Secondly,  we hope that the results in this section could serve  as a  guide to set up a numerical bootstrap scheme for higher genus $(N>3)$ or non-minimal theories.

As we already discussed in Sec.~\ref{sec:bound_str_cts}, the combination of the decomposition of Eq.~\eqref{conformal_block_decomposition_0}
and the crossing symmetry condition in Eq.~\eqref{cross_symmetry} leads to Eq.~\eqref{cross_symmetry_1}. 
If we now normalize to one the structure constant of the channel $h_1=h_2=\cdots=h_N=0$, that is $D_{\boldsymbol{0}, \boldsymbol{0}}=1$, then Eq.~\eqref{cross_symmetry_1} can be rewritten as
\begin{multline}\label{cross_symmetry_2}
\sum_{\substack{\boldsymbol{h},\boldsymbol{\bar{h}} \\ (\boldsymbol{h},\boldsymbol{\bar{h}})\neq(\boldsymbol{0},\boldsymbol{0})}}D_{\boldsymbol{h},\boldsymbol{\bar{h}}}
 \left[\mathcal{G}_{c,\boldsymbol{h}}^{(N)}(z)\mathcal{G}_{c, \boldsymbol{\bar{h}}}^{(N)}(\bar{z})-
 \mathcal{G}_{c,\boldsymbol{h}}^{(N)}(1-z)\mathcal{G}_{c,\boldsymbol{\bar{h}}}^{(N)}(1-\bar{z})\right]
 \\=\mathcal{G}_{c,\boldsymbol{0}}^{(N)}(1-z)\mathcal{G}_{c, \boldsymbol{0}}^{(N)}(1-\bar{z})-
 \mathcal{G}_{c,\boldsymbol{0}}^{(N)}(z)\mathcal{G}_{c,\boldsymbol{0}}^{(N)}(\bar{z}).
\end{multline}
For the minimal models  introduced in Sec.~\ref{sec:null_vectors}, 
the number of channels in Eq.~\eqref{cross_symmetry_2} is finite. For instance, for $N=2$, it corresponds to the number of conformal families of $\mathcal{C}$ and for $N=3$, is the number of 
non-zero structure constants. Therefore, for a given point $z$, the crossing symmetry relation 
is a linear equation in  $N_c$ (the number of channels except $(\boldsymbol{0}, \boldsymbol{0})$) 
unknowns $D_{\boldsymbol{h},\boldsymbol{\bar{h}}}$. Note that this procedure boils down to choose a different linear functional to be applied to the crossing equation than the one used in Eq.~\eqref{linfunder}. 

As done in ~\cite{SR}, we can draw uniformly $N_c$ random points $\{z_{j}\}$ in the square $[1/2-\kappa, 1/2+\kappa]\times[-i\kappa, i\kappa]$.  Let us also require that each point is separated from the real axis and each other point by a distance 
\begin{equation}
 \delta=\frac{\kappa}{\sqrt{N_c}}
\end{equation}
where $\kappa$ is an arbitrary positive number (which we will fix later). By imposing Eq.~\eqref{cross_symmetry_2} at each $z=z_j$, one obtains a linear system with $N_c$ unknowns and $N_c$ equations. By truncating the expansion in Eq.~\eqref{usap} of the conformal blocks $\mathcal{G}_{c, \boldsymbol{h}}^{(N)}(z)$
at given $L, L'$, it is possible to calculate a set of structure constants $D_{\boldsymbol{h},\boldsymbol{\bar{h}}}(L, L')$ for any random realization of the points $\{z_j\}$. If the bootstrap converges,  we expect that the variance of the solutions $D_{\boldsymbol{h},\boldsymbol{\bar{h}}}(L, L')$ will be small and that $D_{\boldsymbol{h},\boldsymbol{\bar{h}}}(L, L')\to D_{\boldsymbol{h},\boldsymbol{\bar{h}}}$, for $L, L'\to \infty$.  

As a first benchmark of the method, we have considered the Tricritical 
Ising CFT with $c=7/10$ on a torus ($N=2$). 
In Appendix~\ref{app_min_mod}, we gather the operator content
of this model. It belongs to the diagonal series of the minimal models, namely  its partition function $\mathcal{Z}_1$ on the torus, see Eq.~\eqref{pf}, is 
\begin{equation}
\mathcal{Z}_1(x)=\sum_{h}|\chi_{c,h}(\tau(x))|^2,
\end{equation}
for $c=7/10$ and the sum running over the conformal dimensions in Eq.~\eqref{tric_ising_fields}. We first used the regularization prescription explained in Section \ref{null_vec1} to compute the combinatorial expansion of Eq.~\eqref{genus_one_conf_block} until level $L=6$. From it, we obtained the approximation for the $N=2$ orbifold conformal blocks of Eq.~\eqref{usap}, truncated at $L,L'=6$.
In Table~\ref{tab_tric_results}, we have considered 100 different sets of
random points $\{z_j\}$ and we have calculated the mean and 
the coefficient of  variance (the standard deviation divided by the mean) 
of the values for the structure constants derived by solving Eq.~\eqref{cross_symmetry_2} 
for each sample of random points. One can see that the numerical bootstrap converges.

\begin{table}[tbp]
\centering
\begin{tabular}{|c|c|c|}
\hline 
  \multirow{2}{*}{$h$} &  \multicolumn{2}{c|}{$D_{\boldsymbol{h},\boldsymbol{h}}$} \\
  \cline{2-3}
                       & numerical bootstrap  & exact (Eq.~\eqref{D_2})  \\
 \hline
 $\sigma$        &  $0.65975046\,(8.8\times 10^{-5})$ & $0.65975396$\\
 $\sigma'$       &  $0.00781211\,(8.0\times 10^{-4})$ & $0.00781250$\\
 $\varepsilon$   &  $0.32987828\,(6.4\times 10^{-5})$ & $0.32987698$\\
 $\varepsilon'$  &  $0.00128874\,(1.9\times 10^{-3})$ & $0.00128858$\\
 $\varepsilon''$ &  $5.92566\times10^{-8}\,(2.7\times 10^{-1})$ & $5.96046\times 10^{-8}$ \\
 \hline
\end{tabular}
\caption{\label{tab_tric_results} Results of the numerical bootstrap for the Tricritical 
Ising CFT  on the torus $(N=2)$. The first column indicates the channel labelled by the
corresponding primary field, see Eq.~\eqref{tric_ising_fields}. The second column corresponds to the mean value for 
each structure constant calculated after considering 100 different sets of random points with $\kappa=0.22$. 
We have truncated the expansion in sphere conformal blocks and the elliptic recursion at level $L,L'=6$. In brackets, the 
coefficient of variation.}
\end{table}

We have  then applied the numerical bootstrap to determine the $N=3$ twist field correlation function for the following  models: The Ising CFT ($c=1/2$), the  Lee-Yang CFT ($c=-22/5$), 
and the Gaffnian CFT ($c=-3/5$) \cite{Simon,Ardonne}. Note that the last two theories are non-unitary, see Refs.~\cite{Dupic, CastroAlv} for related studies. All of them 
belong to the diagonal series of minimal models and, therefore, only pairings between holomorphic and antiholomorphic primaries with the same conformal dimensions ($\boldsymbol{h}=\boldsymbol{\bar{h}}$) are possible. From Eq.~\eqref{D_3} and Eq.~\eqref{cross_symmetry_2} we can extract numerically the seed CFT structure constant as $(C^{\text{seed}}_{\boldsymbol{h},\boldsymbol{h}})^2=27^{2|\boldsymbol{h}|}\;D_{\boldsymbol{h}, \boldsymbol{h}}$. In Appendix~\ref{app_min_mod}, we remind the field content and the fusion rules of the minimal
models under consideration.

Table~\ref{tab_bootstrap_results} shows the results of the $N=3$ numerical bootstrap for these three
CFTs: Ising  (top), Lee-Yang (center) and the Gaffnian  (bottom). 
We consider again 100 different sets $\{z_j\}$ of random points. In the second column of each chart, we write 
the mean and the coefficient of variance computed for  $(C^{\text{seed}}_{\boldsymbol{h},\boldsymbol{h}})^2$   after solving 
Eq.~\eqref{cross_symmetry_2} with each set of random points. The third column collects 
the exact results for  
the square of the seed structure constants~\cite{DF}. The agreement between the two values is excellent.

\begin{table}[tbp]
\centering 

\begin{tabular}{|c|c|c|}
\multicolumn{3}{c}{$c=1/2$}\\
\hline 
  \multirow{2}{*}{$(h_1, h_2, h_3)$} &  \multicolumn{2}{c|}{$(C^{\text{seed}}_{\boldsymbol{h},\boldsymbol{h}})^2$} \\
  \cline{2-3}
                                     &  numerical bootstrap & exact \\ 
 \hline
 $(\sigma, \sigma, \mathbb{I})$ & $1.00001359\, (9.7\times 10^{-5})$  &    1   \\
 $(\varepsilon, \varepsilon, \mathbb{I})$ & $1.00032363\, (2.8\times 10^{-3})$  &    1   \\
 $(\sigma, \sigma, \varepsilon)$   & $0.24993371\, (2.0\times 10^{-3})$ & 0.25 \\
 \hline
\end{tabular}
\vspace{0.75cm}

\begin{tabular}{|c|c|c|}
\multicolumn{3}{c}{$c=-22/5$}\\
\hline 
  \multirow{2}{*}{$(h_1, h_2, h_3)$} &  \multicolumn{2}{c|}{$(C^{\text{seed}}_{\boldsymbol{h},\boldsymbol{h}})^2$} \\
  \cline{2-3} 
                                     & numerical bootstrap & exact\\
 \hline
 $(\varphi, \varphi, \mathbb{I})$ &  $\phantom{-}0.99999804\, (8.9\times 10^{-6})$ &   1   \\
 $(\varphi, \varphi, \varphi)$ &    $-3.65310941\, (8.5\times 10^{-6})$ &  $-3.65311624$  \\
 \hline
\end{tabular}
\vspace{0.75cm}

\begin{tabular}{|c|c|c|}
\multicolumn{3}{c}{$c=-3/5$}\\
\hline 
  \multirow{2}{*}{$(h_1, h_2, h_3)$} &  \multicolumn{2}{c|}{$(C^{\text{seed}}_{\boldsymbol{h},\boldsymbol{h}})^2$} \\
  \cline{2-3} 
                                     & numerical bootstrap & exact\\
 \hline
 $(\sigma, \sigma, \mathbb{I})$ &            $\phantom{-}0.99981087\,(4.1\times10^{-4})$ & 1        \\
 $(\varepsilon, \varepsilon, \mathbb{I})$ &  $\phantom{-}0.99983570\,(1.0\times10^{-3})$ & 1        \\
 $(\psi, \psi, \mathbb{I})$   &              $\phantom{-}0.99629757\,(2.1\times10^{-2})$ & 1 \\
 $(\sigma, \sigma, \varepsilon)$ &           $-0.27381762\ (4.1\times 10^{-4})$          &$-0.27373889$\\
 $(\varepsilon, \varepsilon, \varepsilon)$ & $-4.37665844\, (2.2\times 10^{-3})$         &$-4.37982231$\\
 $(\sigma, \varepsilon, \psi)$ &             $\phantom{-}0.24859929\,(1.1\times 10^{-2})$ & 0.25\\
 \hline
\end{tabular}

\caption{\label{tab_bootstrap_results} Results of the numerical bootstrap for 
the Ising (top), Lee-Yang  (center), and the Gaffnian (bottom)  CFTs
on a genus two $\mathbb Z_3$-symmetric Riemann surface ($N=3$). For each minimal model, the second column 
collects the mean value determined for each structure constant after performing the bootstrap
with 100 different sets of random points $\{z_j\}$, fixing $\kappa=0.22$. The sphere conformal block
expansion and the elliptic recursion were truncated at level $L, L'=6$. In the brackets we write the coefficient of variation.
The third column contains the exact value of $(C_{\boldsymbol{h}, \boldsymbol{h}}^{\text{seed}})^2$, calculated in~\cite{DF} for the minimal models.}
\end{table}

\begin{figure}[t]
 \begin{minipage}{0.5\linewidth}
 \centering 
 \includegraphics[width=\textwidth]{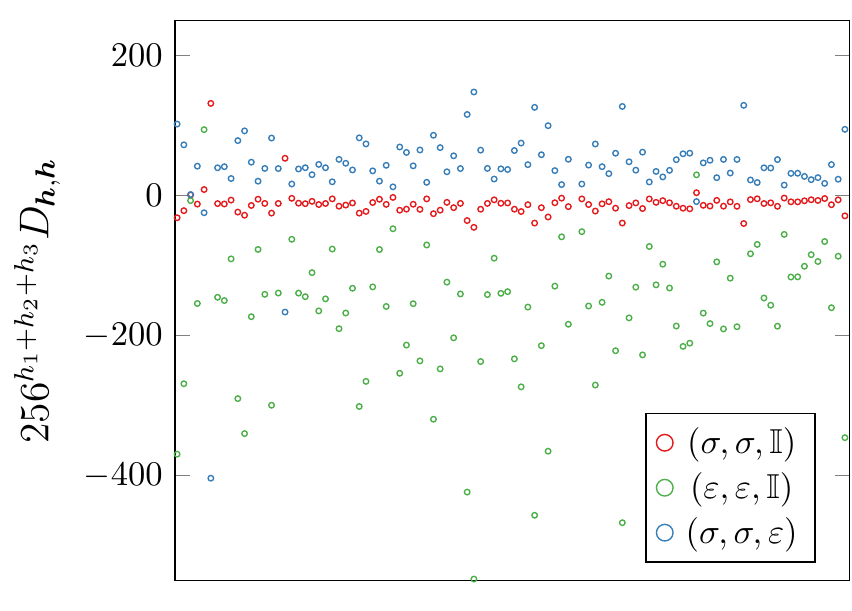}
\end{minipage}
 \begin{minipage}{0.5\linewidth}
 \centering 
 \includegraphics[width=0.93\textwidth]{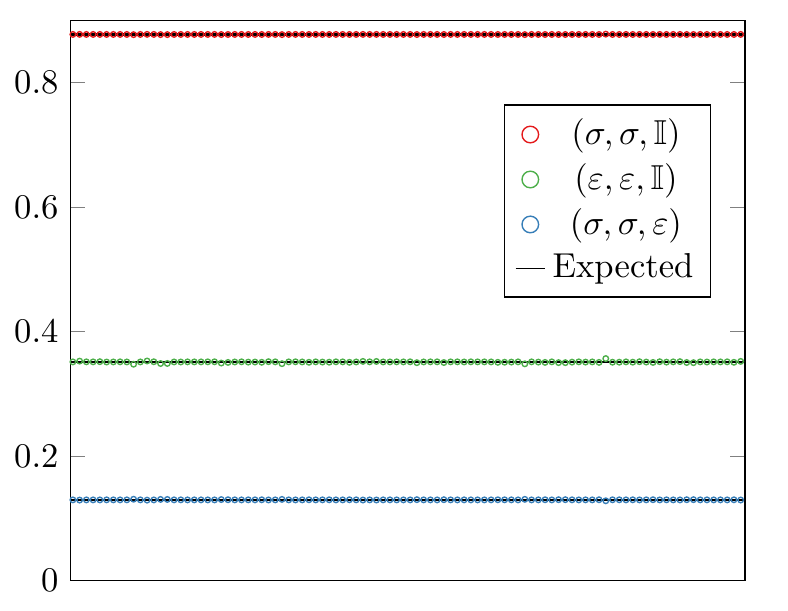}
\end{minipage}
\caption{{\label{comparison}}We plot the values obtained for the $N=3$ structure constants $D_{\boldsymbol{h},\boldsymbol{h}}$ 
(rescaled by a factor for representation purposes) in the Ising CFT by solving the cross-symmetry Eq.~\eqref{cross_symmetry_2} for 100 different
sets of random points, choosing $\kappa=0.22$, and considering two different approximations for the 
orbifold conformal blocks. On the left, we have taken the small $z$ expansion derived for them in 
Eq.~\eqref{genus_two_conf_block}, truncated at order $L=6$. On the right, we have 
expanded them in sphere conformal blocks, as in Eq.~\eqref{expansion_sphere_conf_blocks_trunc}, until $L=6$ and we have applied
the elliptic recursion of Eq.~\eqref{zam_recursion} to compute the latter up to level $L'=6$. The solid lines correspond to the 
values expected for the structure constants.}
\end{figure}

Finally, in Figure \ref{comparison}, we compare the numerical solutions of Eq.~\eqref{cross_symmetry_2}  obtained by first replacing the orbifold conformal blocks with their power series expansion about $z=0$ (on the left) and  then by their representation in terms of sphere conformal blocks (on the right).  The results are shown for the Ising CFT. One can see that in the first case the bootstrap approach cannot predict the correct structure constants. This ultimately can be traced back to the $z\mapsto 1-z$ asymmetry that is visible in Figure \ref{fig:ent_ising}.

\section{Conclusions}

In this paper we analyzed  conformal four-point twist field correlation functions in the $\mathbb Z_N$ orbifold. Such correlators can be also interpreted as CFT partition functions on Riemann surfaces of genus $N-1$, with  a cyclic symmetry. We focused in particular on seed theories $\mathcal{C}$ which belong to the series of the minimal models with $c<1$. From a CFT perspective, we extended the results in~\cite{Collier} in two directions. Firstly, we provided a regularization scheme for the combinatorial expansion of the orbifold conformal blocks that gets rid of  all the null vectors when $c<1$. Then, we  proposed a method to systematically expand the genus $N-1$ conformal blocks into sphere conformal blocks of central charge $Nc$. The latter are more suitable for applications, since they can be calculated, by the recursion formula~\cite{Zamolodchikov}, as power series in the elliptic variable $q$.

The sphere conformal blocks, which appear in the decomposition, can be identified order by order in $q$ only by comparison with the combinatorial expansion of the orbifold conformal blocks that incorporates the descendant states in the replicated theory $\mathcal{C}^{\otimes N}$. The inclusion of the descendant contributions, and their evaluation through conformal mappings, represents a substantial improvement to the previous power series in $q$  of the twist field correlators~\cite{Rajabpour, Ruggiero}.
 
We examined in detail the case $N=3$, i.e. genus two Riemann surfaces, and discussed extensive applications of the formalism. In particular, we calculated R\'enyi entropies for two disjoint intervals in minimal CFTs, reproducing with great accuracy the available results for the free Majorana fermion and providing new ones for interacting theories.  We also showed how the partially transposed reduced density matrix for two disjoint intervals could be calculated in a conformal block expansion and applied this result to the Tricritical Ising CFT. The representation in terms of  sphere conformal blocks of the twist field correlators was also employed to refine the bound on the structure constants of unitary theories found in~\cite{Collier}.

 There are a couple of possible future directions that are worth to be mentioned. First, it would be useful to investigate whether our formalism can be extended  to arbitrary values of the genus $g=N-1$. This task involves the determination of the descendant $N$-point function in Eq.~\eqref{N-point} for a rational CFT.  We expect the latter to be analytic in $g$  at least for a free compactified boson, thus allowing to recover the results for the entanglement entropy and negativity discussed in~\cite{ Furukawa, Alba3, Calabrese09, CalabreseNeg, Grava} from a different route.
Also it will be important to understand if a recursive formula for higher genus conformal blocks, such as that put forward in~\cite{Cho}, can be effectively implemented when $c<1$, due to the additional null vector resonances.

\acknowledgments
We thank Andrea Cappelli, Nina Javerzat, Sylvain Ribault and Erik Tonni for stimulating discussions. We are indebted to Benoit Estienne for important comments on a first version of the paper. FA and JV somehow acknowledge partial support by the Brazilian Ministries MEC and MCTC, the CNPq (grant number 306209/2019-5), the Simons Foundation (Grant Number 884966, AF) and the Italian Ministry MIUR under the grant PRIN 2017  ``Low-dimensional quantum systems: theory, experiments and simulations". RS thanks the IIP, Natal, where this project started. FA is grateful to Maurizio Fagotti for his kind invitation to LPTMS, Paris.  Finally, we acknowledge Institut Pascal, Paris, and the program ``Bootstat 2021'' for hospitality during the final stages of this work. 
\appendix
\section{Transformation Properties of Virasoro Descendants}\label{app_ward}

Consider a CFT on the compactified complex plane (Riemann sphere) with coordinate $z$. The Virasoro 
generators are defined by their action on the fields,
\begin{equation}
\label{vir_ln}
 L_{-n}(z)\phi^{M}_h(z)=\oint_{C_{z}}\frac{dz'}{2\pi i}(z'-z)^{-n+1}~T(z')~\phi^{M}_h(z),
\end{equation}
where $C_{z}$ is a closed contour containing $z$ and $T(z)$ is the stress energy tensor. 
From the current-current OPE,
\begin{equation}
T(z) T(0)= \frac{c/2}{z^4}+ \frac{2}{z^2} T(0)+\frac{1}{z}\partial T (0)+\text{regular terms},
\end{equation} 
one obtains the  Virasoro algebra
\begin{equation}
 [L_n(0), L_{m}(0)]\phi^{M}_h(0) =\left[(n-m) L_{n+m}(0)+\frac{c}{12}n(n^2-1)\delta_{n+m, 0}\right]\phi^{M}_h(0).
\end{equation}

The Virasoro generators acting on a field inserted at $z=\infty$ are given by
\begin{equation}\label{L_infty}
 L_{-n}(z=\infty)=-\oint_{C_{\infty}}\frac{dz}{2\pi i}z^{n+1}T(z).
\end{equation}
Let us see how these operators transform under a conformal map
$t \mapsto z(t)$. By recalling the transformation of the stress energy tensor 
when we apply it,
\begin{equation}
T(z)\mapsto \left(\frac{d z}{d t}\right)^{-2} \left[T(t)-\frac{c}{12}\left\{z(t),t\right\}\right],
\end{equation}
where $\{z(t),t\}$ is the Schwarzian derivative, the Virasoro descendant in Eq.~\eqref{L_infty}
is then transformed into the linear combination of descendants acting on the point
$t_\infty$ in the $t$-surface whose image is $z(t_\infty)=\infty$
\begin{equation}\label{Virasoro_transf}
 \mathcal{L}_{-n}(t_\infty)=-\oint_{C_{t_\infty}}\frac{dt}{2\pi i}
 \left(\frac{dz}{dt}\right)^{-1}[z(t)]^{n+1}\left[T(t)-\frac{c}{12}\{z(t), t\}\right],
\end{equation}
where $C_{t_\infty}$ is a contour encircling the point $t_\infty$, and $T(t)$ can be 
expressed as
\begin{equation}
 T(t)=\sum_{m\in\mathbb{Z}}(t-t_\infty)^{-m-2}L_m(t=t_\infty).
\end{equation}

The results in Sec.~\ref{sec:genus_one} can be derived by evaluating Eq.~\eqref{Virasoro_transf} with the map given in Eq.~\eqref{g_one_covering_map}.
In particular, we can  write down the expansion of the Virasoro descendants in Eq.~\eqref{Virasoro_transf}
by applying the residue theorem; if $n\geq 1$ one has
\begin{equation}\label{vir_g1_1}
 \mathcal{L}_{-n}(t_{\infty})=\sum_{m\geq -n} a_{nm} L_{m}(t=t_{\infty})+\frac{c}{32}(n-1), \quad t_{\infty}=\{0,\infty\},
\end{equation}
with
\begin{equation}\label{vir_g1_2}
 a_{nm}=\frac{1}{4^n}\left[2^{2n+1}-\binom{2n+1}{m+n+1}{}_2F_1(1, m-n, m+n+2, -1)\right].
\end{equation}

In the case $N=3$, studied in Sec.~\ref{sec:genus_two}, the orbifold three-point function of Eq.~\eqref{C_3_0} is calculated by
applying the conformal map of Eq.~\eqref{g_two_covering_map}. The pullbacks $\mathcal{L}_{-n}(t_\infty)$ of the Virasoro descendants 
$L_{-n}(z=\infty)$ under this map can be also derived by using Eq.~\eqref{Virasoro_transf}. We have, for $n\geq 1$,
\begin{equation}\label{vir_g2} 
 \mathcal{L}_{-n}(t_\infty)=\sum_{m\geq n}a_{nm}^{t_\infty}L_{m}(t=t_{\infty})+\frac{c}{27}(n-1),
 \quad t_\infty=\{0, 1, \infty\}
\end{equation}
where the coefficients $a_{nm}^{t_\infty}$ can be determined in closed form from the residue theorem, see also Eq.~(2.9) of Ref.~\cite{Collier}.

The three-point correlations on the sphere that appear in the expansion in Eq.~\eqref{genus_two_conf_block} of the $N=3$ orbifold 
conformal block can be computed recursively by employing the following Ward identities~\cite{Teschner},
\begin{multline}
\label{wardvir1}
 \langle L_{-n}\phi_{h_1}^{M_1}|\phi_{h_2}^{M_2}(1)|\phi_{h_3}^{M_3}\rangle=
 \langle \phi_{h_1}^{M_1}|\phi_{h_2}^{M_2}(1)|L_n \phi_{h_3}^{M_3}\rangle \\
 +\sum_{m= 1}^n \binom{n+1}{m+1}\langle \phi_{h_1}^{M_1}|L_m \phi_{h_2}^{M_2}(1)|\phi_{h_3}^{M_3}\rangle,
\end{multline}
\begin{multline}
\label{wardvir2}
 \langle \phi_{h_1}^{M_1}|L_{-n}\phi_{h_2}^{M_2}(1)|\phi_{h_3}^{M_3}\rangle=
 \sum_{m=0}^\infty\binom{n+m-2}{m}
 \left[\langle L_{m+n} \phi_{h_1}^{M_1}|\phi_{h_2}^{M_2}(1)|\phi_{h_3}^{M_3}\rangle \right.\\ +
 \left. (-1)^n\langle \phi_{h_1}^{M_1}|\phi_{h_2}^{M_2}(1)|L_{m-1} \phi_{h_3}^{M_3}\rangle\right],
\end{multline}
and
\begin{equation}
 \langle \phi_{h_1}|\phi_{h_2}(1)|L_{-n}\phi_{h_3}\rangle=
 \langle L_{-n}\phi_{h_3}|\phi_{h_2}(1)|\phi_{h_1}\rangle.
\end{equation}

\section{Orbifold Virasoro Sub-algebra}\label{app_orbifold_algebra}

In each sheet of the orbifold theory $\mathcal{C}^{\otimes N}/\mathbb{Z}_N$, we 
can consider a copy $\boldsymbol{T}^{(j)}(z)$, $j=1,\dots, N$, of the stress-energy tensor of 
the seed theory $\mathcal{C}$. Then the stress-energy tensor of the 
orbifold theory $\boldsymbol{T}(z)$ is 
\begin{equation}\label{hatT}
 \boldsymbol{T}(z)=\sum_{j=1}^N \boldsymbol{T}^{(j)}(z).
\end{equation}
It generates transformations affecting all the sheets in the same way. 
Its Fourier modes,
\begin{equation}
 \boldsymbol{L}_n=\oint_{C_0}\frac{dz}{2\pi i} z^{n+1}\boldsymbol{T}(z),
\end{equation}
where $C_0$ is a contour encircling the point $z=0$, are symmetric 
under the exchange of sheets, 
\begin{equation}
 \boldsymbol{L}_n=\sum_{j=1}^N \boldsymbol{L}_n^{(j)}, \quad 
 \boldsymbol{L}_n^{(j)}=\mathbb{I}\otimes \overset{j}{\cdots} \otimes L_{n}\otimes \cdots \otimes \mathbb{I},
\end{equation}
and form a Virasoro algebra, 
\begin{equation}\label{symm_virasoro_alg}
 [\boldsymbol{L}_n, \boldsymbol{L}_{m}]=(n-m)\boldsymbol{L}_{n+m}+\frac{Nc}{12}n(n^2-1)\delta_{n+m, 0},
\end{equation}
with central charge $Nc$.

\section{Entanglement Entropy and Logarithmic Negativity}\label{app_ent}

Let us consider a generic quantum system that can be divided into two 
spatial regions, which we call $A$ and $B$, such that the total Hilbert 
space factorizes as $\mathcal{H}=\mathcal{H}_A\otimes \mathcal{H}_B$.
We suppose that the system is in a pure state $|\Psi\rangle\in\mathcal{H}$. 
Hence the state of subsystem $A$ is described by the reduced density matrix
\begin{equation}\label{red_density_matrix}
\rho_A=\Tr_{\mathcal{H}_B}|\Psi\rangle\langle\Psi|,
\end{equation}
with $\Tr_{\mathcal{H}_B}$ denoting the partial trace in the space $\mathcal{H}_B$. 
The entanglement between regions $A$ and $B$ can be analysed using the moments of the 
reduced density matrix, $\Tr\rho_A^N$. In particular, the entanglement entropy 
\begin{equation}
 S_A=-\Tr(\rho_A\log \rho_A)
\end{equation}
can be calculated from the R\'enyi entanglement entropies 
\begin{equation}
 S_A^{(N)}=\frac{1}{1-N}\log\Tr\rho_A^N
\end{equation}
by exploiting the so-called  replica trick~\cite{Holzhey, CalCardy04}
\begin{equation}
 S_A=\lim_{N\to 1^+}S_A^{(N)}=-\lim_{N\to1^+}\frac{\partial}{\partial N}\Tr\rho_A^N.
\end{equation}
If subsystem $A$ consists of two disjoint regions $A_1$ and $A_2$, 
such that and $\mathcal{H}_A=\mathcal{H}_{A_1}\otimes\mathcal{H}_{A_2}$,
then one can consider the partial transpose $\rho_A^{T_2}$ of $\rho_A$. 
If $\{|e_j^{(l)}\rangle\}$ denotes a basis of the space $\mathcal{H}_{A_l}$, 
then the matrix elements of $\rho_A^{T_2}$ are defined as 
\begin{equation}\label{partial_transpose}
\langle e_{j_1}^{(1)}e_{k_1}^{(2)}| \rho_A^{T_2} |e_{j_2}^{(1)}e_{k_2}^{(2)}\rangle=
\langle e_{j_1}^{(1)}e_{k_2}^{(2)}| \rho_A |e_{j_2}^{(1)}e_{k_1}^{(2)}\rangle.
\end{equation}
The moments $\Tr(\rho_A^{T_2})^N$ encode the entanglement between regions $A_1$ and $A_2$. 
A particular measure of the entanglement between these two regions is for instance the 
logarithmic negativity~\cite{Vidal},
\begin{equation}
 \mathcal{E}=\log\Tr|\rho_A^{T_2}|,
\end{equation}
which can also be calculated through the replica trick~\cite{CalabreseNeg12, CalabreseNeg}
\begin{equation}
 \mathcal{E}=\lim_{n_{\rm e}\to 1^+}\log\Tr(\rho_A^{T_2})^{n_{\rm e}},
\end{equation}
by taking the analytic continuation of the moments $\Tr(\rho_A^{T_2})^N$
with even exponent $N=n_{\rm e}$.

\section{Operator Content and Fusion Rules in the Minimal Models considered}\label{app_min_mod}

In this Appendix, we recollect the list of primary fields and fusion rules for the 
conformal minimal models under consideration in the paper. The general expression 
for the central charge $c_{p, q}$ and the conformal dimension of the primaries of 
these models are respectively given in Eqs.~\eqref{central_charge_min_mod} and \eqref{conf_dim_min_mod}.

\begin{itemize}
 \item Lee-Yang CFT: $(p, q)=(5,2)$, $c_{5, 2}=-22/5$
       \begin{equation}
 \begin{tabular}{cc}
 \begin{tabular}{|c|c|}
  \hline
  $\phi_h$        & $h$ \\
  \hline
  $\varphi$  & $-1/5$    \\
  \hline
 \end{tabular}
 &
 \hskip 0.5cm 
 \begin{tabular}{l}
  $\varphi \times \varphi = \mathbb{I} +\varphi$ \\
 \end{tabular}
 \end{tabular}
\end{equation} 
\item Ising CFT: $(p, q)=(4,3)$, $c_{4, 3}=1/2$
\begin{equation}\label{ising_fields}
\begin{tabular}{cc}
 \begin{tabular}{|c|c|}
  \hline
  $\phi_h$        & $h$ \\
  \hline
  $\sigma$  & $1/16$    \\
  $\varepsilon$ & $1/2$ \\
  \hline
 \end{tabular}
 &
 \hskip 0.5cm 
 \begin{tabular}{l}
  $\sigma \times \sigma = \mathbb{I} +\varepsilon$ \\
  $\varepsilon \times \varepsilon = \mathbb{I}$ \\
  $\sigma \times \varepsilon = \sigma$\\
 \end{tabular}
\end{tabular}
\end{equation}

\item Gaffnian CFT: $(p, q)=(5, 3)$, $c_{5, 3}=-3/5$
\begin{equation}
  \begin{tabular}{cc}
 \begin{tabular}{|c|c|}
  \hline
  $\phi_h$        & $h$ \\
  \hline
  $\sigma$  & $-1/20$    \\
  $\psi$     & $3/4$      \\
  $\varepsilon$ & $1/5$  \\
  \hline
 \end{tabular}
 &
 \hskip 0.5cm 
 \begin{tabular}{lcl}
  $\sigma \times \sigma = \mathbb{I} +\varepsilon$ & & $\sigma \times \varepsilon = \sigma +\psi$ \\
  $\varepsilon\times \varepsilon = \mathbb{I} + \varepsilon$ & & $\sigma \times \psi = \varepsilon$\\
  $\psi \times \psi = \mathbb{I}$ & &  $\varepsilon \times \psi = \sigma$\\
 \end{tabular}
 \end{tabular}
\end{equation}

\item Tricritical Ising CFT: $(p, q)=(5, 4)$, $c_{5, 4}=7/10$

\begin{equation}\label{tric_ising_fields}
  \begin{tabular}{cc}
 \begin{tabular}{|c|c|}
  \hline
  $\phi_h$        & $h$ \\
  \hline
  $\sigma$  & $3/80$    \\
  $\sigma'$     & $7/16$      \\
  $\varepsilon$ & $1/10$  \\
  $\varepsilon'$ &  $3/5$     \\
  $\varepsilon''$ & $3/2$     \\
  \hline
 \end{tabular}
 &
 \hskip 0.5cm 
 \begin{tabular}{lcl}
  $\varepsilon \times \varepsilon = \mathbb{I} +\varepsilon'$ & & $\sigma \times \sigma = \mathbb{I} +\varepsilon+\varepsilon'+\varepsilon''$ \\
  $\varepsilon'\times \varepsilon' = \mathbb{I} + \varepsilon'$ & & $\sigma' \times \sigma' = \mathbb{I}+\varepsilon''$\\
  $\varepsilon \times \varepsilon' = \varepsilon+\varepsilon''$ & &  $\sigma \times \sigma' = \varepsilon+\varepsilon'$\\
  $\varepsilon \times \sigma' = \sigma$                         & &  $\varepsilon'\times \sigma'= \sigma$ \\
  $\varepsilon \times \sigma = \sigma'+\sigma$                  & &  $\varepsilon'\times \sigma = \sigma'+\sigma$ \\
 \end{tabular}
 \end{tabular}
\end{equation}

\end{itemize}

\section{$N=3$ Orbifold Conformal Blocks for the Ising CFT}\label{app_ising}

For the Ising CFT, according to its fusion rules in Eq.~\eqref{ising_fields}, the conformal block decomposition of 
Eq.~\eqref{conformal_block_decomposition_0} of the $\mathbb{Z}_3$ twist field four-point function takes the form
\begin{eqnarray}\label{orbifold_cb_ising}
 \langle \sigma_3(\infty)\tilde{\sigma}_3(1)\sigma_3(z, \bar{z})\tilde{\sigma}_3(0)\rangle &=&
 \Big|\mathcal{G}_{\frac{1}{2},\{0, 0, 0\}}^{(3)}(z)\Big|^2+
 3D_{\sigma\sigma\mathbb{I}}\Big|\mathcal{G}_{\frac{1}{2},\{\frac{1}{16},\frac{1}{16}, 0\}}^{(3)}(z)\Big|^2 \nonumber\\ & & +
 3D_{\varepsilon\varepsilon\mathbb{I}}\Big|\mathcal{G}_{\frac{1}{2},\{\frac{1}{2}, \frac{1}{2}, 0\}}^{(3)}(z)\Big|^2
 +3D_{\sigma\sigma\varepsilon}\Big|\mathcal{G}_{\frac{1}{2},\{\frac{1}{16}, \frac{1}{16}, \frac{1}{2}\}}^{(3)}(z)\Big|^2,
\end{eqnarray}
where 
\begin{equation}
 D_{\sigma\sigma\mathbb{I}}=\frac{1}{3^{3/4}},\quad
 D_{\varepsilon\varepsilon\mathbb{I}}=\frac{1}{729},\quad
 D_{\sigma\sigma\varepsilon}=\frac{1}{4\cdot 3^{15/4}}. 
\end{equation}

The $N=3$ orbifold conformal blocks above have the following 
expansions in terms of Virasoro sphere conformal blocks up to level $L=5$,
cf. Eq.~\eqref{expansion_sphere_conf_blocks_trunc},
\begin{equation}
 \mathcal{G}_{\frac{1}{2},\{0, 0, 0\}}^{(3)}(z)\sim
 \mathcal{F}_{\frac{3}{2}, 0}(z) + 
 \frac{49}{10451673} \mathcal{F}_{\frac{3}{2}, 4}(z) + 
 \frac{2}{4782969} \mathcal{F}_{\frac{3}{2}, 5}(z),
\end{equation}

\begin{multline}
 \mathcal{G}_{\frac{1}{2},\{\frac{1}{16}, \frac{1}{16}, 0\}}^{(3)}(z)\sim
 \mathcal{F}_{\frac{3}{2},\frac{1}{8}}(z) + 
 \frac{1}{432} \mathcal{F}_{\frac{3}{2}, \frac{9}{8}}(z) + 
 \frac{2209}{2612736} \mathcal{F}_{\frac{3}{2}, \frac{17}{8}}(z) + 
 \frac{590597}{5965996032} \mathcal{F}_{\frac{3}{2}, \frac{25}{8}}(z) \\ + 
 \frac{61593283775}{9375929753665536}\mathcal{F}_{\frac{3}{2}, \frac{33}{8}}(z)  + 
 \frac{13237693484267}{24583253711470460928}\mathcal{F}_{\frac{3}{2}, \frac{41}{8}}(z),
\end{multline}

\begin{multline}
 \mathcal{G}_{\frac{1}{2},\{\frac{1}{2},\frac{1}{2},0\}}^{(3)}(z)\sim
 \mathcal{F}_{\frac{3}{2}, 1}(z)+ 
 \frac{1}{54} \mathcal{F}_{\frac{3}{2}, 2}(z)+ 
 \frac{1}{3402}\mathcal{F}_{\frac{3}{2}, 3}(z) + 
 \frac{2401}{8109396}\mathcal{F}_{\frac{3}{2}, 4}(z) \\+ 
 \frac{6245}{1499726502}\mathcal{F}_{\frac{3}{2}, 5}(z) + 
 \frac{539}{12033950004}\mathcal{F}_{\frac{3}{2},6}(z), 
\end{multline}
and
\begin{multline}
 \mathcal{G}_{\frac{1}{2},\{\frac{1}{16},\frac{1}{16},\frac{1}{2}\}}^{(3)}(z)\sim
 \mathcal{F}_{\frac{3}{2}, \frac{5}{8}}(z) + 
 \frac{49}{432}\mathcal{F}_{\frac{3}{2}, \frac{13}{8}}(z) + 
 \frac{637}{124416}\mathcal{F}_{\frac{3}{2}, \frac{21}{8}}(z) + 
 \frac{176647}{1988665344}\mathcal{F}_{\frac{3}{2}, \frac{29}{8}}(z) \\ + 
 \frac{6395744863}{1086928370270208}\mathcal{F}_{\frac{3}{2}, \frac{37}{8}}(z) + 
 \frac{528656973059}{299112879787868160}\mathcal{F}_{\frac{3}{2}, \frac{45}{8}}(z). 
\end{multline}

For the Ising CFT, the function $\mathcal{R}_N(z)$ that appears in Eqs.~\eqref{F_n_cc}
and $\eqref{G_n_cc}$ is of the form~\cite{Calabrese11, Tagliacozzo}
\begin{equation}\label{R_n_ising}
\mathcal{R}_N^{\text{Ising}}(z)=\frac{1}{2^{N-1}|\chartheta{\boldsymbol{0}}{\boldsymbol{0}}(\Omega(z))|}
\sum_{\boldsymbol{\varepsilon},\boldsymbol{\delta}}
\left|\chartheta{\boldsymbol{\varepsilon}}{\boldsymbol{\delta}} (\Omega(z))\right|,\quad z\in\mathbb{C},
\end{equation}
where
$$\chartheta{\boldsymbol{\varepsilon}}{\boldsymbol{\delta}} (\Omega)=
\sum_{\boldsymbol{m}\in\mathbb{Z}^{N-1}}
e^{i\pi(\boldsymbol{m}+\boldsymbol{\varepsilon})^t
\cdot \Omega(\boldsymbol{m}+\boldsymbol{\varepsilon})+2\pi i 
(\boldsymbol{m}+\boldsymbol{\varepsilon})^t\cdot \boldsymbol{\delta}}.
$$
The characteristics of the Theta function above are half-integer vectors, $\boldsymbol{\varepsilon},\boldsymbol{\delta}\in(\mathbb{Z}/2)^{N-1}$. The sum in Eq.~\eqref{R_n_ising} runs over all the vectors $\boldsymbol{\varepsilon}$, $\boldsymbol{\delta}$ with components $\varepsilon_j, \delta_j\in\{0, 1/2\}$, and $\Omega(z)$ is the symmetric $(N-1)\times (N-1)$ matrix 
\begin{equation}\label{period_matrix}
\Omega_{rs}(z)=\frac{2 i}{N}\sum_{k=1}^{N-1}\sin\left(\frac{\pi k}{N}\right)
\cos\left[\frac{2\pi k}{N}(r-s)\right]\beta_{k/N}(z),
\end{equation}
in which
$$\beta_{k/N}(z)=\frac{_2F_1(k/N, 1-k/N, 1; 1-z)}{_2F_1(k/N, 1-k/N, 1; z)}.$$

\section{$N=3$ Orbifold Conformal Blocks for the Tricritical Ising CFT}\label{app_tric_ising}
If we take into account the fusion rules of Eq.~\eqref{tric_ising_fields}, the decomposition of the 
$\mathbb{Z}_3$ twist field four-point correlation function in the Tricritical Ising CFT reads
\begin{eqnarray}
 \langle \sigma_3(\infty)\tilde{\sigma}_3(1)\sigma_3(z, \bar{z})\tilde{\sigma}_3(0)\rangle & =&
 \Big|\mathcal{G}_{\frac{7}{10},\{0,0,0\}}^{(3)}(z)\Big|^2+
 3D_{\sigma\sigma\mathbb{I}}\Big|\mathcal{G}_{\frac{7}{10},\{\frac{3}{80},\frac{3}{80},0\}}^{(3)}(z)\Big|^2 \nonumber \\ & &+
 3D_{\sigma'\sigma'\mathbb{I}}\Big|\mathcal{G}_{\frac{7}{10},\{\frac{7}{16},\frac{7}{16},0\}}^{(3)}(z)\Big|^2+
 3D_{\varepsilon\varepsilon\mathbb{I}}\Big|\mathcal{G}_{\frac{7}{10}, \{\frac{1}{10}, \frac{1}{10}, 0\}}^{(3)}(z)\Big|^2 \nonumber \\ & &+
 3D_{\varepsilon'\varepsilon'\mathbb{I}}\Big|\mathcal{G}_{\frac{7}{10}, \{\frac{3}{5},\frac{3}{5},0\}}^{(3)}(z)\Big|^2+
 3D_{\varepsilon''\varepsilon''\mathbb{I}}\Big|\mathcal{G}_{\frac{7}{10},\{\frac{3}{2},\frac{3}{2},0\}}^{(3)}(z)\Big|^2 \nonumber \\ & &+
 3D_{\sigma\sigma\varepsilon}\Big|\mathcal{G}_{\frac{7}{10},\{\frac{3}{80},\frac{3}{80},\frac{1}{10}\}}^{(3)}(z)\Big|^2+
 3D_{\sigma\sigma\varepsilon'}\Big|\mathcal{G}_{\frac{7}{10},\{\frac{3}{80},\frac{3}{80},\frac{3}{5}\}}^{(3)}(z)\Big|^2 \nonumber \\ & &+
 3D_{\sigma\sigma\varepsilon''}\Big|\mathcal{G}_{\frac{7}{10},\{\frac{3}{80},\frac{3}{80},\frac{3}{2}\}}^{(3)}(z)\Big|^2 +
 6D_{\sigma\sigma'\varepsilon}\Big|\mathcal{G}_{\frac{7}{10}, \{\frac{3}{80}, \frac{7}{16}, \frac{1}{10}\}}^{(3)}(z)\Big|^2 \nonumber \\ & &+
 6D_{\sigma\sigma'\varepsilon'}\Big|\mathcal{G}_{\frac{7}{10}, \{\frac{3}{80}, \frac{7}{16}, \frac{3}{5}\}}^{(3)}(z)\Big|^2 +
 3D_{\sigma'\sigma'\varepsilon''}\Big|\mathcal{G}_{\frac{7}{10}, \{\frac{7}{16}, \frac{7}{16}, \frac{3}{2}\}}^{(3)}(z)\Big|^2 \nonumber \\ & & +
 3D_{\varepsilon\varepsilon\varepsilon'}\Big|\mathcal{G}_{\frac{7}{10}, \{\frac{1}{10},\frac{1}{10},\frac{3}{5}\}}^{(3)}(z)\Big|^2+
 6D_{\varepsilon\varepsilon'\varepsilon''}\Big|\mathcal{G}_{\frac{7}{10}, \{\frac{1}{10}, \frac{3}{5}, \frac{3}{2}\}}^{(3)}(z)\Big|^2 \nonumber \\ & & +
 D_{\varepsilon'\varepsilon'\varepsilon'}\Big|\mathcal{G}_{\frac{7}{10}, \{\frac{3}{5}, \frac{3}{5}, \frac{3}{5}\}}^{(3)}(z)\Big|^2
\end{eqnarray}
The structure constants $D_{h_1,h_2,h_3}$ above are determined, according to Eq.~\eqref{D_3},
by the OPE coefficients of the Tricritical Ising CFT, whose analytic values can be found for instance
in Ref.~\cite{Mussardo}.

The expansions in sphere conformal blocks until level $L=5$ of the $N=3$ orbifold conformal blocks 
that appear in the previous expression read
\begin{equation}
  \mathcal{G}_{\frac{7}{10},\{0,0,0\}}^{(3)}(z)\sim
  \mathcal{F}_{\frac{21}{10}, 0}(z) +
  \frac{17}{3838185}\mathcal{F}_{\frac{21}{10}, 4}(z) +
  \frac{2}{4782969}\mathcal{F}_{\frac{21}{10}, 5}(z),
\end{equation}

\begin{multline}
\mathcal{G}_{\frac{7}{10},\{\frac{3}{80},\frac{3}{80},0\}}^{(3)}(z)\sim 
\mathcal{F}_{\frac{21}{10},\frac{3}{40}}(z)+\frac{1}{720}\mathcal{F}_{\frac{21}{10},\frac{43}{40}}(z)+
\frac{3265267}{2547417600}\mathcal{F}_{\frac{21}{10}, \frac{83}{40}}(z)+
\frac{3665197}{50512896000}\mathcal{F}_{\frac{21}{10}, \frac{123}{40}}(z)\\
+\frac{155865980497261283}{18705929999865937920000}\mathcal{F}_{\frac{21}{10},\frac{163}{40}}(z) 
+\frac{462436566571594180410731}{659386782637573125832704000000}\mathcal{F}_{\frac{21}{10},\frac{203}{40}}(z),
\end{multline}

\begin{multline}
 \mathcal{G}_{\frac{7}{10},\{\frac{7}{16},\frac{7}{16},0\}}^{(3)}(z)\sim
 \mathcal{F}_{\frac{21}{10}, \frac{7}{8}}(z)+
 \frac{7}{432}\mathcal{F}_{\frac{21}{10}, \frac{15}{8}}(z)+
 \frac{125}{6594048}\mathcal{F}_{\frac{21}{10}, \frac{23}{8}}(z)+
 \frac{7557625}{34022301696} \mathcal{F}_{\frac{21}{10}, \frac{23}{8}}(z) \\
 +\frac{136592048239}{41822362293239808} \mathcal{F}_{\frac{21}{10}, \frac{39}{8}}(z)
 +\frac{3691268343947}{18937730481342382080} \mathcal{F}_{\frac{21}{10},\frac{47}{8}}(z),
\end{multline}

\begin{multline}
 \mathcal{G}_{\frac{7}{10}, \{\frac{1}{10},\frac{1}{10}, 0\}}^{(3)}(z)\sim 
 \mathcal{F}_{\frac{21}{10},\frac{1}{5}}(z) + 
 \frac{1}{270} \mathcal{F}_{\frac{21}{10},\frac{6}{5}}(z) + 
 \frac{4489}{6718950} \mathcal{F}_{\frac{21}{10}, \frac{11}{5}}(z) +
 \frac{734651}{6957940500} \mathcal{F}_{\frac{21}{10},\frac{16}{5}}(z) \\ +
 \frac{33710613028}{7001301644623125} \mathcal{F}_{\frac{21}{10},\frac{21}{5}}(z)+
 \frac{561570243431821}{1229479741472989781250} \mathcal{F}_{\frac{21}{10}, \frac{26}{5}}(z),
\end{multline}

\begin{multline}
  \mathcal{G}_{\frac{7}{10}, \{\frac{3}{5}, \frac{3}{5}, 0\}}^{(3)}\sim
  \mathcal{F}_{\frac{21}{10}, \frac{6}{5}}(z) +
  \frac{1}{45} \mathcal{F}_{\frac{21}{10}, \frac{11}{5}}(z) +
  \frac{12664}{12885075} \mathcal{F}_{\frac{21}{10}, \frac{16}{5}}(z) +
  \frac{75304}{304266375} \mathcal{F}_{\frac{21}{10},\frac{21}{5}}(z) \\ +
  \frac{5224025025746}{910725734424436875} \mathcal{F}_{\frac{21}{10},\frac{26}{5}}(z) +
  \frac{41871765632}{94232186890734375} \mathcal{F}_{\frac{21}{10},\frac{31}{5}}(z),  
\end{multline}

\begin{multline}
  \mathcal{G}_{\frac{7}{10},\{\frac{3}{2}, \frac{3}{2}, 0\}}^{(3)}(z)\sim
  \mathcal{F}_{\frac{21}{10}, 3}(z) +
  \frac{1}{18} \mathcal{F}_{\frac{21}{10}, 4}(z) +
  \frac{230}{22113}\mathcal{F}_{\frac{21}{10}, 5}(z) +
  \frac{100}{443961}\mathcal{F}_{\frac{21}{10}, 6}(z) \\ +
  \frac{96002975}{10064284515558}\mathcal{F}_{\frac{21}{10}, 7}(z) +
  \frac{58456305937}{17581104541057428}\mathcal{F}_{\frac{21}{10},8}(z),
\end{multline}

\begin{multline}
  \mathcal{G}_{\frac{7}{10}, \{\frac{3}{80}, \frac{3}{80}, \frac{1}{10}\}}^{(3)}(z)\sim
  \mathcal{F}_{\frac{21}{10}, \frac{7}{40}}(z) +
  \frac{5}{1296}\mathcal{F}_{\frac{21}{10}, \frac{47}{40}}(z) +
  \frac{475}{373248} \mathcal{F}_{\frac{21}{10}, \frac{87}{40}}(z) +
  \frac{14170687}{97122115584} \mathcal{F}_{\frac{21}{10}, \frac{127}{40}}(z) \\ +
  \frac{71253834712735}{4997056637667115008}\mathcal{F}_{\frac{21}{10}, \frac{167}{40}}(z)+
  \frac{86091919475921327}{95239084395063869964288}\mathcal{F}_{\frac{21}{10}, \frac{207}{40}}(z),
\end{multline}

\begin{multline}
  \mathcal{G}_{\frac{7}{10}, \{\frac{3}{80}, \frac{3}{80}, \frac{3}{5}\}}^{(3)}(z)\sim
  \mathcal{F}_{\frac{21}{10}, \frac{27}{40}}(z) +
  \frac{5}{16} \mathcal{F}_{\frac{21}{10}, \frac{67}{40}}(z) +
  \frac{2476157}{118070784} \mathcal{F}_{\frac{21}{10}, \frac{107}{40}}(z)+
  \frac{555083831}{1041791901696}  \mathcal{F}_{\frac{21}{10}, \frac{147}{40}}(z) \\ +
  \frac{35552447512576825}{476153434940080914432} \mathcal{F}_{\frac{21}{10}, \frac{187}{40}}(z)+
  \frac{152443525738528231}{16655797786474708992000} \mathcal{F}_{\frac{21}{10}, \frac{227}{40}}(z),
\end{multline}

\begin{multline}
  \mathcal{G}_{\frac{7}{10}, \{\frac{3}{80}, \frac{3}{80}, \frac{3}{2}\}}^{(3)}(z)\sim
  \mathcal{F}_{\frac{21}{10}, \frac{63}{40}}(z) +
  \frac{169}{80} \mathcal{F}_{\frac{21}{10}, \frac{103}{40}}(z) +
  \frac{4956437831}{16444684800} \mathcal{F}_{\frac{21}{10}, \frac{143}{40}}(z) +
  \frac{40879507267951}{2130203225088000}\mathcal{F}_{\frac{21}{10}, \frac{183}{40}}(z) \\ +
  \frac{48838420003812575817893}{41322611540300358942720000} \mathcal{F}_{\frac{21}{10}, \frac{223}{40}}(z) +
  \frac{853096471556262783844477}{9087626895163066810368000000}\mathcal{F}_{\frac{21}{10}, \frac{263}{40}}(z),
\end{multline}

\begin{multline}
  \mathcal{G}_{\frac{7}{10},\{\frac{3}{80}, \frac{7}{16}, \frac{1}{10}\}}^{(3)}(z)\sim
  \mathcal{F}_{\frac{21}{10}, \frac{23}{40}}(z) +
  \frac{1301}{15120} \mathcal{F}_{\frac{21}{10}, \frac{63}{40}}(z) +
  \frac{3195223}{653990400} \mathcal{F}_{\frac{21}{10}, \frac{103}{40}}(z)+
  \frac{111395056711}{745930902528000} \mathcal{F}_{\frac{21}{10}, \frac{143}{40}}(z) \\ +
  \frac{69649784481358065419}{11050452065922572943360000}\mathcal{F}_{\frac{21}{10}, \frac{183}{40}}(z) + 
  \frac{4967477167662960708160673}{3123989432446707136069632000000} \mathcal{F}_{\frac{21}{10}, \frac{223}{40}}(z),
\end{multline}

\begin{multline}
  \mathcal{G}_{\frac{7}{10}, \{\frac{3}{80}, \frac{7}{16}, \frac{3}{5}\}}^{(3)}(z)\sim
  \mathcal{F}_{\frac{21}{10}, \frac{43}{40}}(z) +
  \frac{1423}{45360} \mathcal{F}_{\frac{21}{10}, \frac{83}{40}}(z) +
  \frac{237113}{210470400} \mathcal{F}_{\frac{21}{10}, \frac{123}{40}}(z) +
  \frac{3550509597637}{9426871673856000} \mathcal{F}_{\frac{21}{10}, \frac{163}{40}}(z) \\ +
  \frac{134673192184141631833}{6783814636189023928320000} \mathcal{F}_{\frac{21}{10}, \frac{203}{40}}(z) +
  \frac{17146721811911780522868659}{15403567241995839166611456000000} \mathcal{F}_{\frac{21}{10}, \frac{243}{40}}(z),
\end{multline}

\begin{multline}
  \mathcal{G}_{\frac{7}{10}, \{\frac{7}{16}, \frac{7}{16}, \frac{3}{2}\}}^{(3)}(z)\sim
  \mathcal{F}_{\frac{21}{10}, \frac{19}{8}}(z) +
  \frac{289}{3024} \mathcal{F}_{\frac{21}{10}, \frac{27}{8}}(z) +
  \frac{113339}{101380608} \mathcal{F}_{\frac{21}{10}, \frac{35}{8}}(z) +
  \frac{795728725}{5786102439936} \mathcal{F}_{\frac{21}{10}, \frac{43}{8}}(z) \\ +
  \frac{36813418825}{914736180363264} \mathcal{F}_{\frac{21}{10}, \frac{51}{8}}(z) +
  \frac{5346115131945399913}{1885729815637402501251072} \mathcal{F}_{\frac{21}{10}, \frac{59}{8}}(z),
\end{multline}

\begin{multline}
  \mathcal{G}_{\frac{7}{10}, \{\frac{1}{10}, \frac{1}{10}, \frac{3}{5}\}}^{(3)}(z)\sim
  \mathcal{F}_{\frac{21}{10}, \frac{4}{5}}(z) +
  \frac{5}{54} \mathcal{F}_{\frac{21}{10}, \frac{9}{5}}(z) +
  \frac{25}{9828}  \mathcal{F}_{\frac{21}{10}, \frac{14}{5}}(z) +
  \frac{733}{9893988} \mathcal{F}_{\frac{21}{10}, \frac{19}{5}}(z) + \\
  \frac{27915235}{2181879918072} \mathcal{F}_{\frac{21}{10}, \frac{24}{5}}(z) +
  \frac{178775929427}{123348562272120978}  \mathcal{F}_{\frac{21}{10}, \frac{29}{5}}(z),
\end{multline}

\begin{multline}
  \mathcal{G}_{\frac{7}{10}, \{\frac{1}{10}, \frac{3}{5}, \frac{3}{2}\}}^{(3)}(z)\sim
  \mathcal{F}_{\frac{21}{10}, \frac{11}{5}}(z) +
  \frac{173}{810} \mathcal{F}_{\frac{21}{10}, \frac{16}{5}}(z) +
  \frac{753719}{114945075} \mathcal{F}_{\frac{21}{10}, \frac{21}{5}}(z) +
  \frac{108761449}{205479038250}  \mathcal{F}_{\frac{21}{10}, \frac{26}{5}}(z) \\ +
  \frac{409782125417}{6102655912923750} \mathcal{F}_{\frac{21}{10}, \frac{31}{5}}(z) +
  \frac{105239895304351091}{24755684834277326812500} \mathcal{F}_{\frac{21}{10}, \frac{36}{5}}(z),
\end{multline}
and
\begin{multline}
  \mathcal{G}_{\frac{7}{10}, \{\frac{3}{5}, \frac{3}{5}, \frac{3}{5}\}}^{(3)}(z)\sim
  \mathcal{F}_{\frac{21}{10}, \frac{9}{5}}(z) +
  \frac{961}{458055} \mathcal{F}_{\frac{21}{10}, \frac{19}{5}}(z) +
  \frac{355}{341172} \mathcal{F}_{\frac{21}{10}, \frac{24}{5}}(z) + 
  \frac{46601998346}{761410878222969} \mathcal{F}_{\frac{21}{10}, \frac{29}{5}}(z)\\ +
  \frac{819218}{544871045511} \mathcal{F}_{\frac{21}{10}, \frac{34}{5}}(z).
\end{multline}



\begin{thebibliography}{99}

\bibitem{BPZ} A. Belavin, A. Polyakov, A. B. Zamolodchikov. \emph{Infinite conformal symmetry in two-dimensional quantum field theory},
\href{https://doi.org/10.1016/0550-3213(84)90052-X}{\emph{Nucl. Phys. B}, {\bf 241} (1984) 333-380}.

\bibitem{Ribault} S. Ribault, \emph{Conformal field theory on the plane}, \href{https://arxiv.org/abs/1406.4290}{\texttt{arXiv:1406.4290 [hep-th]}}.

\bibitem{ZZ} A. B. Zamolodchikov, Al. B. Zamolodchikov, \emph{Structure Constants and Conformal Bootstrap in Liouville Field Theory},
\href{https://doi.org/10.1016/0550-3213(96)00351-3}{\emph{Nucl. Phys. B} {\bf 477}, 577 (1996)}.

\bibitem{DV} G. Delfino, J. Viti, \emph{On three-point connectivity in two-dimensional percolation},
\href{https://doi.org/10.1088/1751-8113/44/3/032001}{\emph{J. Phys. A: Math. Theor.} {\bf 44} (2011) 032001}.

\bibitem{Picco} M. Picco, S. Ribault, R. Santachiara, \emph{On four-point connectivities in the critical 2d Potts model},
\href{https://scipost.org/10.21468/SciPostPhys.7.4.044}{\emph{SciPost Phys.} {\bf 7}, 044 (2019)}.

\bibitem{He} Y. He, J. L. Jacobsen, H. Saleur, \emph{Geometrical four-point functions in the two-dimensional critical $Q$-state Potts model: The interchiral conformal bootstrap}, 
\href{https://doi.org/10.1007/JHEP12(2020)019}{\emph{JHEP} {\bf 12} (2020) 019}.

\bibitem{Ludwig} A. W. W. Ludwig, \emph{Critical Behavior of the Two-dimensional Random $Q$ State Potts
Model by Expansion in $(Q-2)$}, 
\href{https://doi.org/10.1016/0550-3213(87)90330-0}{\emph{Nucl. Phys. B} {\bf 285}, 97 (1987)}.

\bibitem{CardyBook} J. Cardy, \emph{Scaling and Renormalization in Statistical Physics}, Cambridge
University Press (1996).

\bibitem{DotsenkoJacobsen} V. Dotsenko, J. L. Jacobsen, M.-A. Lewis, M. Picco, \emph{Coupled Potts models: Self-duality and fixed point structure}, 
\href{https://doi.org/10.1016/S0550-3213(99)00097-8}{\emph{Nucl. Phys. B} {\bf 546} [FS], 505 (1999)}.

\bibitem{Komargodski} Z. Komargodski, D. Simmons-Duffin, \emph{The Random-Bond Ising Model in 2.01 and 3 Dimensions},
\href{https://doi.org/10.1088/1751-8121/aa6087}{\emph{J. Phys. A: Math. Theor.} 50 154001 (2017)}.

\bibitem{DelfinoRev} G. Delfino, \emph{Particles, conformal invariance and criticality in pure and disordered systems},
\href{https://doi.org/10.1140/epjb/s10051-021-00076-0}{\emph{Eur. Phys. J. B} {\bf 94} (2021) 65}.

\bibitem{Holzhey} C. Holzhey, F. Larsen, F. Wilczek, \emph{Geometric and Renormalized Entropy in Conformal Field Theory},
\href{https://doi.org/10.1016/0550-3213(94)90402-2}{\emph{Nucl. Phys. B} {\bf 424}, 443 (1994)}.

\bibitem{CalCardy04} P. Calabrese, J. Cardy, \emph{Entanglement Entropy and Quantum Field Theory}, 
\href{https://doi.org/10.1088/1742-5468/2004/06/P06002}{\emph{J. Stat. Mech.} (2004) P06002}.

\bibitem{Caraglio} M. Caraglio, F. Gliozzi, \emph{Entanglement Entropy and Twist Fields},
\href{https://doi.org/10.1088/1126-6708/2008/11/076}{\emph{JHEP} {\bf 11} (2008) 076}.

\bibitem{Calabrese09} P. Calabrese, J. Cardy, E. Tonni, \emph{Entanglement entropy of two disjoint intervals in conformal field theory},
\href{https://doi.org/10.1088/1742-5468/2009/11/P11001}{\emph{J. Stat. Mech.} (2009) P11001}.

\bibitem{Calabrese11} P. Calabrese, J. Cardy, E. Tonni, \emph{Entanglement entropy of two disjoint intervals in conformal field theory II}, \href{https://doi.org/10.1088/1742-5468/2011/01/P01021}{\emph{J. Stat. Mech.} P01021 (2011)}.

\bibitem{Vidal} G. Vidal, R. F. Werner, \emph{Computable measure of entanglement},
\href{https://doi.org/10.1103/PhysRevA.65.032314}{\emph{Phys. Rev. A} {\bf 65}, 032314 (2002)}.

\bibitem{CalabreseNeg12} P. Calabrese, J. Cardy, E. Tonni, \emph{Entanglement Negativity in Quantum Field Theory},
\href{https://doi.org/10.1103/PhysRevLett.109.130502}{\emph{Phys. Rev. Lett.} {\bf 109}, 130502 (2012)}.

\bibitem{CalabreseNeg} P. Calabrese, J. Cardy, E. Tonni, \emph{Entanglement negativity in extended systems:  A field theoretical approach},
\href{https://doi.org/10.1088/1742-5468/2013/02/P02008}{\emph{J. Stat. Mech.} (2013) P02008}.

\bibitem{Gaberdiel} L. Eberhardt, M. R. Gaberdiel, R. Gopakumar, \emph{Deriving the $\text{AdS}_3/\text{CFT}_2$ correspondence},
\href{https://doi.org/10.1007/JHEP02(2020)136}{\emph{JHEP} {\bf 02} (2020) 136}.

\bibitem{Lunin} O. Lunin, S. D. Mathur, \emph{Correlation functions for $M(N)/S(N)$ orbifolds}, \href{https://doi.org/10.1007/s002200100431}{\emph{Commun. Math. Phys.} {\bf 219} (2001) 399-442}.

\bibitem{Dixon} L. Dixon, D. Friedan, E. Martinec, S. Shenker, \textit{The conformal field theory of orbifolds}, 
\href{https://doi.org/10.1016/0550-3213(87)90676-6}{\emph{Nucl. Phys. B} {\bf 282} (1987) 13-73}.

\bibitem{Knizhnik} V. G. Knizhnik, \emph{Analytic fields on Riemannian surfaces. II}, 
\href{https://doi.org/10.1007/BF01225373}{\emph{Commun. Math. Phys.} {\bf 112}, 567-590 (1987)}.

\bibitem{Headrick} M. Headrick, \emph{Entanglement R\'enyi entropies in holographic theories},
\href{https://doi.org/10.1103/PhysRevD.82.126010}{\emph{Phys. Rev. D} {\bf 82}, 126010 (2010)}.

\bibitem{ZamolodchikovAT} Al. B. Zamolodchikov, \emph{Conformal scalar field on the hyperelliptic curve and critical Ashkin-Teller multipoint correlation functions}, \href{https://doi.org/10.1016/0550-3213(87)90350-6}{\emph{Nucl. Phys. B} {\bf 63} [FS19] (1987) 481-503}.

\bibitem{CardyMod} J. Cardy, \emph{Operator content of two-dimensional conformally invariant theories}, 
\href{https://doi.org/10.1016/0550-3213(86)90552-3}{\emph{Nucl. Phys. B} {\bf 270} [FS16] (1986) 186-204}.

\bibitem{Cardy} J. Cardy, A. Maloney, H. Maxfield, \emph{A new handle on three-point coefficients: OPE asymptotics from genus
two modular invariance}, \href{https://doi.org/10.1007/JHEP10(2017)136}{\emph{JHEP} {\bf 10} (2017) 136}.

\bibitem{Collier} M. Cho, S. Collier, X. Yin, \emph{Genus two modular bootstrap}, \href{https://doi.org/10.1007/JHEP04(2019)022}{\emph{JHEP} {\bf 04} (2019) 22}.

\bibitem{Keller} C. A. Keller, G. Mathys, I. G. Zadeh, \emph{Bootstrapping chiral CFTs at genus two},
\href{https://dx.doi.org/10.4310/ATMP.2018.v22.n6.a3}{\emph{Adv. Theor. Math. Phys.} {\bf 22} (2018) 1447-1487}.

\bibitem{Cho} M. Cho, S. Collier, X. Yin, \emph{Recursive representations of arbitrary Virasoro conformal blocks}, 
\href{https://doi.org/10.1007/JHEP04(2019)018}{\emph{JHEP} {\bf 04} (2019) 018}.

\bibitem{Rajabpour} M. A. Rajabpour, F. Gliozzi, \emph{Entanglement entropy of two disjoint intervals from fusion algebra of twist fields},
\href{https://doi.org/10.1088/1742-5468/2012/02/P02016}{\emph{J. Stat. Mech.} (2012) P02016}.

\bibitem{Ruggiero} P. Ruggiero, P. Calabrese, E. Tonni, \emph{Entanglement entropy of two disjoint intervals and the recursion formula
for conformal blocks}, \href{https://doi.org/10.1088/1742-5468/aae5a8}{\emph{J. Stat. Mech.} (2018) 113101}.

\bibitem{Dupic} T. Dupic, B. Estienne, Y. Ikhlef, \emph{Entanglement entropies of minimal models from null-vectors}, 
\href{https://scipost.org/10.21468/SciPostPhys.4.6.031}{\emph{SciPost Phys.} {\bf 4}, 031 (2018)}.

\bibitem{Zamolodchikov} Al. B. Zamolodchikov, \emph{Conformal symmetry in two-dimensional space: Recursion representation of conformal
block}, \href{https://doi.org/10.1007/BF01022967}{\emph{Theor. Math. Phys.} {\bf 73}, 1088-1093  (1987)}.

\bibitem{Tagliacozzo} P. Calabrese, L. Tagliacozzo, E. Tonni, \emph{Entanglement negativity in the critical Ising chain}, 
\href{https://doi.org/10.1088/1742-5468/2013/05/P05002}{\emph{J. Stat. Mech.} (2013) P05002}.

\bibitem{Alba} V. Alba, \emph{Entanglement negativity and conformal field theory: a Monte Carlo study},
\href{https://doi.org/10.1088/1742-5468/2013/05/P05013}{\emph{J. Stat. Mech.} (2013) P05013}.

\bibitem{Eisler} V. Eisler, Z. Zimboras, \emph{On the partial transpose of fermionic Gaussian states}, 
\href{https://doi.org/10.1088/1367-2630/17/5/053048}{\emph{New J. Phys.} {\bf 17} (2015) 053048}.

\bibitem{Coser} A. Coser, E. Tonni, P. Calabrese, \emph{Partial transpose of two disjoint blocks in XY spin chains},
\href{https://doi.org/10.1088/1742-5468/2015/08/P08005}{\emph{J. Stat. Mech.} (2015) P08005}.

\bibitem{Coser2} A. Coser, E. Tonni, P. Calabrese, \emph{Towards entanglement negativity of two disjoint
intervals for a one dimensional free fermion}, 
\href{https://doi.org/10.1088/1742-5468/2016/03/033116}{\emph{J. Stat. Mech.} (2016) 033116}.

\bibitem{Coser3} A. Coser, E. Tonni, P. Calabrese, \emph{Spin structures and entanglement of two disjoint
intervals in conformal field theories},
\href{https://doi.org/10.1088/1742-5468/2016/05/053109}{\emph{J. Stat. Mech.} (2016) 053109}.

\bibitem{Shapourian} H. Shapourian, P. Ruggiero, S. Ryu, P. Calabrese, \emph{Twisted and untwisted negativity spectrum of free fermions},
\href{https://scipost.org/10.21468/SciPostPhys.7.3.037}{\emph{SciPost Phys.} {\bf 7}, 037 (2019)}.

\bibitem{Grava} T. Grava, A. P. Kels, E. Tonni, \emph{Entanglement of two disjoint intervals in CFT and the 2D Coulomb gas in a lattice},
\href{https://doi.org/10.1103/PhysRevLett.127.141605}{\emph{Phys. Rev. Lett.} {\bf 127}, 141605 (2021)}.

\bibitem{Policastro} M. Kulaxizi, A. Parnachev, G. Policastro, \emph{Conformal Blocks and Negativity at Large Central Charge},
\href{https://doi.org/10.1007/JHEP09(2014)010}{\emph{JHEP} {\bf 09} (2014) 010}.

\bibitem{Ryu} J. Kudler-Flam, S. Ryu, \emph{Entanglement negativity and minimal entanglement wedge cross sections in holographic theories},
\href{https://doi.org/10.1103/PhysRevD.99.106014}{\emph{Phys. Rev. D} {\bf 99}, 106014 (2019)}.

\bibitem{Kusuki} Y. Kusuki, J. Kudler-Flam, S. Ryu, \emph{Derivation of holographic negativity in $\text{AdS}_3/\text{CFT}_2$},
\href{https://doi.org/10.1103/PhysRevLett.123.131603}{\emph{Phys. Rev. Lett.} {\bf 123}, 131603 (2019)}.

\bibitem{Dubrovin} B. A. Dubrovin, A. T. Fomenko, S. P. Novikov, \textit{Modern Geometry---Methods and Applications Part II. The Geometry and Topology of Manifolds}, Springer (1985).

\bibitem{Miranda} R. Miranda, \emph{Algebraic Curves and Riemann Surfaces}, American Mathematical Society (1995).

\bibitem{Whittaker} E.T. Whittaker, G. N. Watson, \emph{A Course in Modern Analysis}, Cambridge University Press (1950).

\bibitem{Cappelli} A. Cappelli, C. Itzykson, J.-B. Zuber, \emph{Modular invariant partition functions in two dimensions}, 
\href{https://doi.org/10.1016/0550-3213(87)90155-6}{\emph{Nucl. Phys.B} {\bf 280} [FS18], 445 (1987)}.

\bibitem{Cappelli2} A. Cappelli, C. Itzykson, J.-B. Zuber, \emph{The A-D-E classification of minimal and $A_1^{(1)}$ conformal invariant theories}, \href{https://doi.org/10.1007/BF01221394}{\emph{Commun. Math. Phys.} {\bf 113}, 1-26 (1987)}.

\bibitem{DiFrancesco} P. Di Francesco, P. Mathieu, D. Senechal, \textit{Conformal field theory}, Springer (1999).

\bibitem{Javerzat} N. Javerzat, R. Santachiara, O. Foda, \emph{Notes on the solutions of Zamolodchikov-type recursion relations in Virasoro minimal models}, \href{https://doi.org/10.1007/JHEP08(2018)183}{\emph{JHEP} {\bf 08} (2018) 183}.

\bibitem{Tanzini} R. Santachiara, A. Tanzini, \emph{Moore-Read fractional quantum Hall wave functions and $SU(2)$ quiver gauge theories}, \href{https://link.aps.org/doi/10.1103/PhysRevD.82.126006}{\emph{Phys.Rev D} {\bf 82} (2010) 126006}.

\bibitem{Alkalaev} K. B. Alkalaev, V. A. Belavin, \emph{Conformal blocks of $\mathcal{W}_N$
minimal models and AGT correspondence}, \href{https://doi.org/10.1007/JHEP07(2014)024}{\emph{JHEP} {\bf 07} (2014) 024}.

\bibitem{FodaAGT} V. Belavin, O. Foda, R. Santachiara, \emph{AGT, $N$-Burge partitions and $\mathcal{W}_N$
minimal models},\href{https://doi.org/10.1007/JHEP10(2015)073}{\emph{JHEP} {\bf 10} (2015) 073}.

\bibitem{SV} R. Santachiara, J. Viti, \emph{Local logarithmic correlators as limits of Coulomb gas integrals},
\href{https://doi.org/10.1016/j.nuclphysb.2014.02.022}{\emph{Nucl. Phys. B} {\bf 882} (2014) 229-262}.

\bibitem{Maldacena} J. Maldacena, D. Simmons-Duffin, A. Zhiboedov, \emph{Looking for a bulk point},
\href{ https://doi.org/10.1007/JHEP01(2017)013}{\emph{JHEP} {\bf 01} (2017) 013}.

\bibitem{Borisov} L. Borisov, M. B. Halpern, C. Schweigert, \emph{Systematic approach to cyclic orbifolds}, 
\href{https://doi.org/10.1142/S0217751X98000044}{\emph{Int. J. Mod. Phys. A} {\bf 13}, 125 (1998)}.

\bibitem{Furukawa}  S. Furukawa, V. Pasquier, J. Shiraishi, \emph{Mutual Information and Boson Radius in $c=1$ Critical Systems in One Dimension}, \href{https://doi.org/10.1103/PhysRevLett.102.170602}{\emph{Phys. Rev. Lett.} {\bf 102}, 170602 (2009)}.

\bibitem{Alba2} V. Alba, L. Tagliacozzo, P. Calabrese, \emph{Entanglement entropy of two disjoint blocks in critical Ising models},
\href{https://doi.org/10.1103/PhysRevB.81.060411}{\emph{Phys. Rev. B} {\bf 81}, 060411(R) (2010)}.

\bibitem{Alba3} V. Alba, L. Tagliacozzo, P. Calabrese, \emph{Entanglement entropy of two disjoint intervals in c=1 theories},
\href{https://doi.org/10.1088/1742-5468/2011/06/P06012}{\emph{J. Stat. Mech.} (2011) P06012}.

\bibitem{Mussardo} G. Mussardo, \emph{Statistical Field Theory: An Introduction to Exactly Solved Models in Statistical Physics}, Oxford University Press (2020).

\bibitem{Feiguin} A. Feiguin, S. Trebst, A. W. W. Ludwig, M. Troyer, A. Kitaev, Z. Wang, M. H. Freedman, \emph{Interacting anyons in topological quantum liquids: The golden chain}, 
\href{https://doi.org/10.1103/PhysRevLett.98.160409}{\emph{Phys. Rev. Lett.} {\bf 98}, 160409 (2007)}.

\bibitem{Rattazzi} R. Rattazzi, V. S. Rychkov, E. Tonni, A. Vichi, \emph{Bounding scalar operator dimensions in 4D CFT},
\href{https://doi.org/10.1088/1126-6708/2008/12/031}{\emph{JHEP} {\bf 12} (2008) 031}.

\bibitem{Poland} D. Poland, S. Rychkov, A. Vichi, \emph{The Conformal Bootstrap: Theory, Numerical Techniques, and Applications},
\href{https://doi.org/10.1103/RevModPhys.91.015002}{\emph{Rev. Mod. Phys.} {\bf 91}, 015002 (2019)}.

\bibitem{SR} S. Ribault, R. Santachiara, \textit{Liouville theory with a central charge less than one}, 
\href{https://doi.org/10.1007/JHEP08(2015)109}{\emph{JHEP} {\bf 08} (2015) 109}.

\bibitem{Simon} S H. Simon, E. H. Rezayi, N. R. Cooper, I. Berdnikov, \emph{Construction of a paired wave function for spinless electrons at filling fraction $\nu=2/5$}, \href{https://doi.org/10.1103/PhysRevB.75.075317}{\emph{Phys. Rev. B} {\bf 75}, 075317 (2007)}.

\bibitem{Ardonne} E. Ardonne, J. Gukelberger, A. W. W. Ludwig, S. Trebst, M. Troyer, \emph{Microscopic models of interacting Yang-Lee anyons},
\href{https://doi.org/10.1088/1367-2630/13/4/045006}{\emph{New J. Phys.} {\bf 13} (2011) 045006}.

\bibitem{CastroAlv} D. Bianchini, O. A. Castro-Alvaredo, B. Doyon, E. Levi, F. Ravanini, \emph{Entanglement entropy of non-unitary conformal field theory},
\href{https://doi.org/10.1088/1751-8113/48/4/04FT01}{\emph{J. Phys. A: Math. Theor.} {\bf 48} 04FT01 (2015)}.

\bibitem{DF} Vl. S. Dotsenko, V. A. Fateev, \emph{Conformal algebra and multipoint correlation functions in 2D statistical models}, \href{https://www.sciencedirect.com/science/article/pii/0550321384902694}{\emph{Nucl. Phys. B} {\bf 284}, 312 (1984)}.

\bibitem{Teschner} J. Teschner, \emph{Liouville theory revisited}, \href{https://doi.org/10.1088/0264-9381/18/23/201}{\emph{Class. Quantum Grav.} {\bf 18} (2001) R153-R222}.



\end{thebibliography}
\end{document}